\documentclass[useAMS,usenatbib,usegraphicx]{mn2e}

\title[Kinematics and Abundances of BCGs and BGGs]{
Spatially resolved kinematics and stellar populations of brightest cluster and group galaxies
}
\author[Brough et al.]{S. Brough$^{1}$\thanks{E-mail:
sbrough@astro.swin.edu.au}, Robert Proctor$^{1}$, Duncan A.~Forbes$^{1}$,  Warrick J.~Couch$^{1}$, C. A.~Collins$^2$, \newauthor D. J.~Burke$^3$, 
R. G.~Mann$^4$
\\$^{1}$Centre for Astrophysics and Supercomputing, Swinburne University of Technology, Hawthorn, VIC 3122, Australia
\\$^2$Astrophysics Research Institute, Liverpool John Moores University, Egerton Wharf, Birkenhead, CH41 1LD, UK
\\$^3$Harvard-Smithsonian Center for Astrophysics, 60 Garden Street, Cambridge, MA 02138, USA
\\$^4$SUPA, Institute for Astronomy, University of Edinburgh, Royal Observatory, Blackford Hill, Edinburgh, EH9 3NJ, UK
}
\begin{document}

\date{Accepted... Received...; in original form 2006}

\pagerange{\pageref{firstpage}--\pageref{lastpage}} \pubyear{2006}

\maketitle

\label{firstpage}

\begin{abstract}
We present an examination of the kinematics and stellar populations of
a sample of 3 Brightest Group Galaxies (BGGs) and 3 Brightest Cluster
Galaxies (BCGs) in X-ray groups and clusters.  We have obtained high
signal-to-noise Gemini/GMOS (Gemini South Multi-Object Spectrograph)
long-slit spectra of these galaxies and use Lick indices to determine
ages, metallicities and $\alpha$-element abundance ratios out to at
least their effective radii.  We find that the BGGs and BCGs have very
uniform masses, central ages and central metallicities.  Examining the
radial dependence of their stellar populations, we find no significant
velocity dispersion, age, or $\alpha$-enhancement gradients.  However,
we find a wide range of metallicity gradients, suggesting a variety of
formation mechanisms.  The range of metallicity gradients observed is
surprising given the homogeneous environment these galaxies probe and
their uniform central stellar populations. However, our results are
inconsistent with any single model of galaxy formation and emphasize
the need for more theoretical understanding of both the origins of
metallicity gradients and galaxy formation itself.  We postulate two
possible physical causes for the different formation mechanisms.
\end{abstract}


\section{Introduction}


Large, bright, elliptical galaxies are found at the centres of most
galaxy clusters.  These brightest cluster galaxies (BCGs) are the most
luminous stellar systems known, yet their intrinsic luminosities are
remarkably uniform.
Their luminosity and photometric uniformity mean that they do not
appear to be the bright extension of the luminosity function of other
cluster galaxies \citep{bernstein01}. The position of these galaxies
at the centre of clusters and their unique properties link their
formation and evolution to that of their environment.  However, the
mechanisms behind their growth are still poorly
understood. Hierarchical models of galaxy formation
(e.g. \citealt{delucia06}; henceforth dLB) suggest that these most
massive galaxies should have assembled more recently than other
galaxies, with those in the densest environments of galaxy clusters
being at a later stage in their evolution to those in the lower
density environments of galaxy groups. Understanding the processes by
which these galaxies form and evolve is, therefore, vital to the
understanding of how both galaxies and clusters form and subsequently
evolve.

The studies of \cite{cm98}, \cite{bcm} and \cite{brough02} compared
the $K$-band properties of BCGs with the X-ray luminosities of their
host clusters. Near-infrared luminosities are dominated by long-lived
stars, such that these wavebands are more sensitive to the underlying
stellar mass of galaxies than shorter wavelengths.  Cluster X-ray
luminosity is directly proportional to the square of the density of
the intra-cluster medium and provides a quantitative and objective
measure of environmental density. \cite{brough02} observed that BCGs
in the most X-ray luminous clusters ($L_{X} > 1.9\times10^{44}$ erg
s$^{-1}$ [$0.3-3.5$ keV]) have uniform absolute magnitudes (after
correction for passive evolution) over redshifts $0.02<z<0.8$.  This
suggests that they have not experienced any significant stellar mass
evolution, beyond that expected for passive evolution, since a
redshift $\sim1$.  In contrast, the absolute magnitudes of BCGs in
less X-ray luminous clusters show significant scatter suggesting that
they have increased their mass by up to a factor $\sim4$ over the same
timescale. An analysis of the structure of the low-redshift ($z<0.1$)
BCGs adds further evidence of environmentally dependent evolution:
BCGs in high-$L_X$ clusters have larger radii with fainter mean
surface brightnesses than those in low-$L_X$ clusters
\citep{brough05}.  This suggests that the BCGs in high-$L_X$ clusters
have undergone more accretion than those in low-$L_X$ clusters. Taken
together with the analysis of the magnitudes, these results suggest
that BCGs in high-mass clusters assembled their stellar mass at
redshifts $>1$ (i.e. $\sim8$ Gyrs ago), and have been passively
evolving since, in contrast to BCGs in low-mass clusters which appear
to still be in the process of assembling today.  This picture is
consistent with models of hierarchical structure formation.

Another important prediction of the hierarchical structure formation
model is that BCGs originate in groups - where they are the brightest
group galaxies (BGGs), generally observed to be large ellipticals
\citep{zabludoff98,osmond04,brough06}.  As the groups fall into
clusters, the BGG sinks to the bottom of the potential well by
dynamical friction and merges with the central galaxy. In
\cite{brough06} we found that, while the luminosity of the BGG
correlates with its total group luminosity, the fraction of group
luminosity contained in the BGG decreases with increasing total group
luminosity. This suggests that BGGs themselves grow in step with their
host group, but at a slower rate than the group.



These observations give indirect evidence for the hierarchical
structure formation paradigm with BCGs in low-mass clusters having
undergone accretion events since $z\sim1$, possibly from mergers with
infalling BGGs.  However, confirming the paradigm requires direct
evidence that BCGs have undergone mergers.
We can use the predictions of dLB to do this in a quantitative
fashion.  Their study uses merger trees from the Millennium
cosmological N-body simulation \citep{springel05} as an input for
their semi-analytic model which includes a model for the suppression
of cooling flows by AGN feedback.  The model is described in more
detail in \cite{delucia06a}.  
dLB predict that the stars that will end up in the BCG form at high
redshift (with 50 per cent of the stars having formed by $z\sim5$,
i.e. they are now $~12.5$ Gyrs old), the BCGs then evolve through the
accretion of old, passive galaxies of mass$~>10^{10} M_{\odot}$, with
$\sim50$ per cent of the mass of the eventual BCG not {\it assembling}
until $z\sim0.5$, i.e. $\sim5$ Gyrs ago.  The late mergers are not
predicted to significantly change the ages or metallicities of the
BCGs because the progenitor galaxies are predicted to be massive, old,
metal-rich, gas-poor galaxies themselves.
dLB also show that the BCG mass grows in step with the growth in mass
of the cluster. We therefore assume that observing BCGs in clusters
with a range in mass (down to group-sized systems) is equivalent to
observing them at different stages in their evolutionary history.

The stellar populations of galaxies provide us with a means by
which to determine their evolutionary histories.  The populations are
a result of many factors including the efficiency of star formation,
gas fraction, gas dissipation and possible interactions with other
galaxies. This information should leave imprints in the radial
distributions of these populations and the different models of galaxy
formation make different predictions of the radial dependence of the
stellar populations:

In classic models of dissipative collapse
(e.g. \citealt{eggen62,larson74,carlberg84,arimoto87}), stars form
during the collapse and remain in their orbits, while gas dissipates
to the centre of the galaxy, being continually enriched by the
evolving stars as it does so.  This results in stars in the centres of
galaxies being more metal-rich than those in the outer regions.  In
the models the degree of dissipation is controlled by the size of the
potential well.  The continuous enrichment by evolving stars means
that the centres are less $\alpha$-element enhanced (i.e. the ratio of
the abundance of $\alpha$-elements to the abundance of iron-peak
elements is $>$ solar; \citealt{tmb}) than the outer regions,
unless the collapse, and star formation, occur over very short
timescales.  The observed central $\alpha$-element enhancement ratios
in galaxies require collapse times $\leq1$ Gyr
\citep{arimoto87,matteucci94}.  More recent numerical models of
dissipative collapse (e.g. \citealt{martinelli98,chiosi02,kawata03})
show that the observed chemical abundances can be obtained by assuming
that metallicity gradients are established by the onset of a galactic
wind which varies depending on the local depth of the potential well.
Dissipative collapse, therefore predicts strong negative metallicity
gradients (i.e. metallicity decreases with increasing radius) that
correlate with galaxy mass \citep{chiosi02,kawata03}, negative
$\alpha$-element enhancement gradients, and small positive to null age
gradients (due to the speed of the collapse).


Simulations of the hierarchical formation of galaxies (e.g. dLB,
\citealt{delucia06a}) predict that massive elliptical galaxies are
produced by mergers.  Early numerical simulations of the properties of
merger remnants suggested that mergers lead to a flattening of the
metallicity gradients \citep{white80}.  However, \cite{vanalbada82}
suggested that the gradients of the progenitor galaxies are not
heavily affected as violent relaxation preserves the position of the
stars in the local potential.  More recent simulations including gas
physics suggest that the gradients of the remnant depend on the amount
of gas present in the merger.  The predicted signatures of gaseous
(dissipative) mergers are qualitatively similar to the dissipative
collapse models described above as a significant fraction of the gas
in the progenitor galaxies is funnelled towards the central regions of
the merger remnant, resulting in a starburst \citep{barnes91}.
Starbursts can also produce metallicity gradients, although those
produced by
\cite{bekki99} in their dissipative merger simulation are shallower than 
those found for dissipative collapse models. \cite{kobayashi04}
simulated the formation of ellipticals in a cold-dark matter
cosmology, examining the effects of dissipative collapse, and
dissipative and dissipationless mergers on the chemodynamical
evolution of ellipticals. She found that galaxies of given mass had
steep metallicity gradients if formed by collapse and shallower
gradients (a few Gyrs after the merger) if formed by a major merger
(dissipative or not).  Therefore, hierarchical galaxy formation models
predict negative metallicity gradients that are shallower than those
predicted by dissipative collapse models.
Depending on the duration and location of the starburst and the
original $\alpha$-element enhancement of the gas, the resulting
$\alpha$-element enhancement ratio and age gradients of the merger
remnant can be either positive or negative (e.g. \citealt{thomas99})

If environment has an effect on the star formation, the number of
interactions a galaxy undergoes, or the dissipation of gas within the
galaxy's potential well, we would expect to see an environmental
dependence on the inferred stellar population gradients.

The theoretical models indicate that the radial structure of a galaxy
provides a fossil record of its evolutionary history. Therefore,
studying the radial kinematics and stellar populations of BGGs and
BCGs and their dependence on environment will enable us to determine
their merger history.
If the predictions of dLB are correct we would expect to see evidence
of dissipationless mergers in these galaxies; i.e. little dependence
of metallicity on system mass and no evidence of star formation in the
last 5 Gyrs.  As the BCGs are predicted to be at a later stage of
their evolution than the BGGs, we would also expect evidence that the
BCGs have undergone more mergers than the BGGs. This would predict,
for example,
shallower metallicity gradients in the BCGs and the velocity
dispersions of the BGGs/BCGs increasing with increasing group/cluster
mass.


There have been few studies of the kinematics and stellar populations
of BCGs and fewer that are spatially resolved to enable a study of
radial trends. \cite{forbes01} found the BCG in the Fornax cluster,
NGC 1399, to be old ($10\pm2$ Gyrs).  
In studies of radial gradients, \cite{gorgas90} found that the Mg$_2$
gradients of 3 BCGs are shallower than the mean gradient of normal
ellipticals, suggesting that BCGs have undergone more mergers than
normal elliptical galaxies.  In contrast, \cite{davidge95} found the
D4000 gradients of 6 BCGs to be steeper than for non-BCGs, suggesting
that BCGs experienced dissipative collapse and cannot have formed
solely through mergers with passive galaxies.  The study of 13 BCGs by
\cite{fisher95a} found that the velocity dispersion profiles of BCGs
were similar to those in normal elliptical galaxies.  
\cite{fisher95b} examined the line strengths of 9 of the 13 BCGs presented 
in \cite{fisher95a} and found that BCGs follow the same Mg$b$ --
velocity dispersion ($\sim$ metallicity -- mass) and [Mg/Fe] --
velocity dispersion ($\sim$ $\alpha$-element enhancement -- mass)
relationships as normal elliptical galaxies.  The H$\beta$ gradients
of their sample are flat, and combined with declining Mg$b$ gradients,
suggest that BCGs are younger in their centres than in their outer
regions. \cite{carter99} found similar results in their sample of 3
BCGs. \cite{mehlert03} calculated stellar population gradients for the
two cD galaxies in the Coma cluster, finding shallow age gradients,
shallow metallicity gradients,
and no $\alpha$-element enhancement ratio gradients, consistent with
their results of other early-type galaxies in Coma.
The results from metallicity gradients are clearly contradictory while
the other relationships are generally consistent with BCGs sharing
similar properties and, therefore, a similar evolutionary history, to
normal early-type galaxies. However, these samples only cover a
limited environmental range and only a few studies have data that
extend to the effective radius ($r_e$; the radius containing half the
light) of these galaxies.



In this paper we test whether BCGs evolve hierarchically
by examining high signal-to-noise, Gemini long-slit spectra for a
sample of 3 BGGs and 3 BCGs in X-ray groups and clusters spanning an
unprecedented range in X-ray luminosity and hence environmental
density ($10^{40}<L_X$ ergs$^{-1}<10^{45}$).  Our aim is not to
compare BGGs with BCGs; instead we are using the range in system X-ray
luminosity to observe these galaxies at different stages in their
evolutionary history.  Observing the same systems over a range of
look-back times would require going to redshifts that are too high for
a detailed stellar population analysis to be conducted.  However, we
do note that it is possible that these galaxies are at different
stages of their evolution and {\it also} in different environments.
If this is so then we might expect the properties of BGGs and BCGs to
show scatter in their relationships and not to depend on their
environment.

We build on previous studies by examining galaxies in a wider range of
environments and by measuring their spectral properties out to at
least 1 effective radius.  We also use the multi-index
$\chi^2$-fitting technique of Proctor \& Sansom (2002; see also
\citealt{proctor04a,proctor04b,proctor05}) to measure their stellar populations.  
The rationale behind this technique is that all the indices defined in
the Lick system have {\it some} sensitivity to both age {\it and}
metallicity. Using them all together, in conjunction with the best
available single stellar population (SSP) models, overcomes
uncertainties in both the data reduction and from the SSP models
themselves.  \cite{proctor04b} demonstrate that this method is
significantly more reliable than the more widely-used technique of
using only single or a few indices at a time.




This is the first in a series of papers studying the kinematics and
stellar populations, and their gradients, in BGGs and BCGs.  At
present we only have a small sample of 6 galaxies.  It is therefore
not possible to determine whether these galaxies follow statistically
significant correlations.  However, the sample is useful for an
initial identification of trends that might be germane to the
aforementioned issues.  

We introduce our sample in Section~\ref{section_sample} and the
observations made and data reduction process in
Section~\ref{section_obs}.  In Section~\ref{section_analysis} we
discuss our method for extracting the kinematics, Lick indices and
ages and metallicities for these data.  In
Section~\ref{section_indiv_gals} we discuss the results we obtain for
individual galaxies before studying the relationships between the
sample in Section~\ref{section_env} (environment),
Section~\ref{section_central} (central stellar populations) and
Section~\ref{section_grads} (stellar population gradients).  We draw
our conclusions in Section~\ref{section_conc}.  Throughout this paper
we assume $H_0=70$ km s$^{-1}$ Mpc$^{-1}$, $\Omega_m=0.3$ and
$\Omega_{\Lambda}=0.7$.


\section{Sample}
\label{section_sample}

The BGG and BCG galaxy samples were selected from the Group Evolution
Multi-wavelength Study (GEMS; \citealt{osmond04,forbes06}) and the BCG
sample presented in \cite{brough05}.  In both these samples the galaxy
is selected as the brightest galaxy
closest to the X-ray centroid, i.e. closest to the centre of the
potential well of these structures.  The specific galaxies were chosen
to be accessible within the Gemini telescope's `A' (February--July)
observing semester and to cover a wide range in group/cluster X-ray
luminosity. The sample of galaxies studied and the properties of their
host group/cluster systems are summarised in Table~\ref{sample_table}.

The absolute magnitudes given in Table~\ref{sample_table} are
calculated from the total $K_s$-band magnitudes (henceforth denoted as
$K$) given in the 2-micron All Sky Survey Extended Source Catalogue
(2MASS; \citealt{jarrett00}).  There are no independent distance
measurements for our BCGs.  We, therefore, convert 
to absolute magnitudes using luminosity distances (given in
Table~\ref{sample_table}) calculated from the measured central
recession velocities and our choice of cosmology .
Due to the scatter in the absolute magnitudes of this sample, it makes
no difference to our results if we use luminosity distances for the
BGGs or those based on distance moduli from surface brightness
fluctuation studies by
\cite{tonry01}.
The magnitudes are then K+E-corrected (i.e. also corrected for passive
evolution; \citealt{yoshii88}), using spectral energy distributions
produced by the GISSEL96 stellar population synthesis code
\citep{bruzual93}, assuming that the galaxies are 10 Gyrs old and 
formed in an instantaneous burst, and have evolved passively since
$z\sim2$.  The correction is only $-0.2$ mag for our most distant
galaxy (2MASX J10172568-1041206 at $z\sim0.054$) and the assumptions
of age and redshift of formation have negligible effect on the
correction value. The magnitudes were also corrected for galactic
absorption using the maps of \cite{schlegel98}, these corrections are
$A_K\sim0.02$ mag.

We take cluster X-ray luminosities from the {\it ROSAT}-ESO
Flux-Limited X-ray galaxy cluster catalogue (REFLEX;
\citealt{bohringer04}).  These are calculated from the {\it ROSAT}
[0.1-2.4 keV] band and extrapolated to 12 times the core radius of the
cluster.  Our group X-ray luminosities are taken from the GEMS group
catalogue \citep{osmond04}.  These luminosities are bolometric and
extrapolated to the radius corresponding to an overdensity of 500
times the critical density -- $r_{500}$.  There is one object in
common between the two samples: the NGC 5044 group.  This group has
Log $L_{X,GEMS}=43.09\pm0.01$ (ergs$^{-1}$) and Log $L_{X,
REFLEX}=43.04$ (ergs$^{-1}$).  REFLEX do not provide errors on their
X-ray luminosities.  Therefore, we assume that the two catalogues are
consistent within errors of $\Delta$Log($L_X)=0.05$ dex.


The 3 BCGs are henceforth referred to by the name of their clusters
i.e. 2MASX J09083238-0937470 -- A754\#1; 2MASX J10172568-1041206 -- A970\#1;
MCG -01-27-002 -- A978\#1.

\begin{table*}
\begin{center}
\caption{A summary of the properties of the sample of galaxies.}
\begin{tabular}{|lcrrccc|}
\hline
Galaxy&System&RA&Dec&Distance&$M_K$&Log ($L_X$)\\
&&(J2000)&(J2000)&(Mpc)&(mag)&(erg s$^{-1}$)\\
\hline
NGC 3557               & NGC 3557  &11:09:57.4&-37:32:17& 44.2&-24.54$\pm$0.01&42.11$\pm0.05$\\
NGC 3640               & NGC 3640  &11:21:06.9& 03:14:06& 19.2&-23.90$\pm$0.02&$<40.74$\\
NGC 5044               & NGC 5044  &13:15:24.0&-16:23:06& 39.6&-25.28$\pm$0.02&43.09$\pm0.05$\\
2MASX J09083238-0937470& Abell 754 &09:08:32.3&-09:37:48&245.1&-26.56$\pm$0.05&44.63$\pm0.05$\\
2MASX J10172568-1041206& Abell 970 &10:17:25.6&-10:41:20&261.3&-25.39$\pm$0.09&44.00$\pm0.05$\\
MCG -01-27-002         & Abell 978 &10:20:26.5&-06:31:36&264.1&-26.50$\pm$0.07&43.43$\pm0.05$\\
\hline
\label{sample_table}
\end{tabular} 
\flushleft The columns indicate (1) galaxy name, (2) Group (NGC) or cluster (Abell) name, (3,4) the position of the galaxy, (5) the luminosity distance to the galaxy, (6) Absolute total $K$-magnitude from 2MASS corrected for galactic extinction and evolution as described in the text, with $1\sigma$ error, (7) System X-ray luminosity and assumed $1\sigma$ error.
\end{center}
\end{table*}

\section{Observations and Data Reduction}
\label{section_obs}

\begin{table*}
\begin{center}
\caption{Observing parameters and relative galaxy sizes.}
\begin{tabular}{|llllllll|}
\hline
Name&Exp. Time&PA&MA&$r_e$&$\epsilon$&$a_e$&Fraction\\ 
&(s)&(degrees)&(degrees)&($^{\prime\prime}$)&&($^{\prime\prime}$)\\
\hline
NGC 3557 &$3\times1200$& 31& 31 &36.7&0.25& 36.7& 1.23  \\
NGC 3640 &$3\times900$&  80& 95 &36.6&0.2 & 25.4& 1.77  \\
NGC 5044 &$5\times1800$& 20& 20 &24.3&0.08& 24.3& 2.06  \\
A754\#1  &$3\times1800$& 96&125&10.0&0.3 &   5.7& 3.68 \\
A970\#1  &$3\times1800$&177& 45&4.4&0.08 &   4.4& 1.60 \\
A978\#1  &$4\times1800$& 18& 10 &8.4&0.2  &  6.6& 2.13\\ 
\hline
\label{obs_table}
\end{tabular}
\flushleft The columns indicate (1) galaxy name, (2) exposure time, (3) 
observed position angle, (4) position angle of major axis from 2MASS,
(5) effective radius of major axis, $r_e$, calculated from 2MASS
$K$-band 20$^{\rm {th}}$ magnitude arcsec$^{-2}$ isophotal radius, (6)
ellipticity of galaxy, from 2MASS, (7) effective radius along observed
position angle, $a_e$, (8) fraction of $a_e$ our radial profiles
extend to.
\end{center}
\end{table*}

We observed the 6 galaxies using the Gemini Multi-Object Spectrograph
(GMOS) on the Gemini (South) telescope in long-slit mode.  All 6
galaxies were observed in queue mode in early 2005; the observation
parameters are summarised in Table~\ref{obs_table}.  The galaxies
cover a range of surface brightnesses and the exposure times reflect
this.  

Table~\ref{obs_table} also gives the position angle (PA) that these
galaxies were observed at, and their respective major axis, from
2MASS.  We note that the spectra of NGC 3557 and NGC 5044 were taken
at the position angles of their major axis whilst the spectra of NGC
3640, A754\#1, A970\#1 and A978\#1 were taken at position angles other
than their major axis, in an attempt to avoid other galaxies lying on
the slit.  This has an effect on the relative sizes of the galaxies at
these position angles.  We therefore also give the effective radius
along the semi-major axis, $r_e$, of the galaxies.  This is calculated
from the 2MASS $K$-band 20$^{\rm {th}}$ magnitude arcsec$^{-2}$
isophotal radius, $r_{K20}$: \cite{jarrett03} show that this is
proportional to $r_e$, as Log $r_e\sim$ Log $r_{K20}-0.4$.  This was
transformed to an effective radius at the PA we observed, $a_e$, using
the ellipticities of the galaxies (also from 2MASS
\citealt{jarrett03}):

\begin{equation}
a_e=\frac{r_e(1-\epsilon)}{1-\epsilon \cos(|PA-MA|)}
\label{rad_eq}
\end{equation}

We also indicate the fraction of $a_e$ our radial profiles span and
note that the profiles of all our galaxies extend beyond $1a_e$.

The GMOS slit is 5.5$^{\prime}$ long and we used a slit width of
$0.5^{\prime\prime}$ with the B600 grating.  The data were binned by 2
in both the spatial and spectral directions.  This set-up gave a
dispersion of 0.9 \AA\ pixel$^{-1}$ and a spectral resolution of
$\sim3.4$\AA\ full width at half-maximum (FWHM; $\sim80$ kms$^{-1}$).
The seeing was generally $\leq1^{\prime\prime}$ and the binned pixel
scale was $0.15^{\prime\prime}$ pixel$^{-1}$. The total wavelength
range covered is $\sim 3700-6700$\AA\ .

We used the GEMINI/GMOS tasks in IRAF for basic data reduction. The
data were then binned in the spatial direction using the
STARLINK/FIGARO package EXTRACT, to guarantee a minimum
signal-to-noise ratio per \AA\ (S/N) of 60 (with the exception of
A970\#1, where only a S/N of 30 was possible) at the wavelength of the
$H\beta$ line.  We also extracted a central aperture of size $a_e/8$
and our slit-width of $0.5^{\prime\prime}$.  Reliable index estimates
were obtained for 22 Lick indices from $H\delta_A$ to Fe5782
\citep{trager98}.

Neither Lick index nor flux calibration standards were observed for
this program. We therefore used standards observed on a different
program using the same GMOS grating in multi-object spectroscopy mode
(\citealt{pierce06}; c.f. \citealt{proctor05}).  These stellar
templates were used to calculate the kinematic properties of the
galaxies.

\begin{figure*}
\begin{center}
     \resizebox{30pc}{!}{
     \rotatebox{-90}{
	\includegraphics{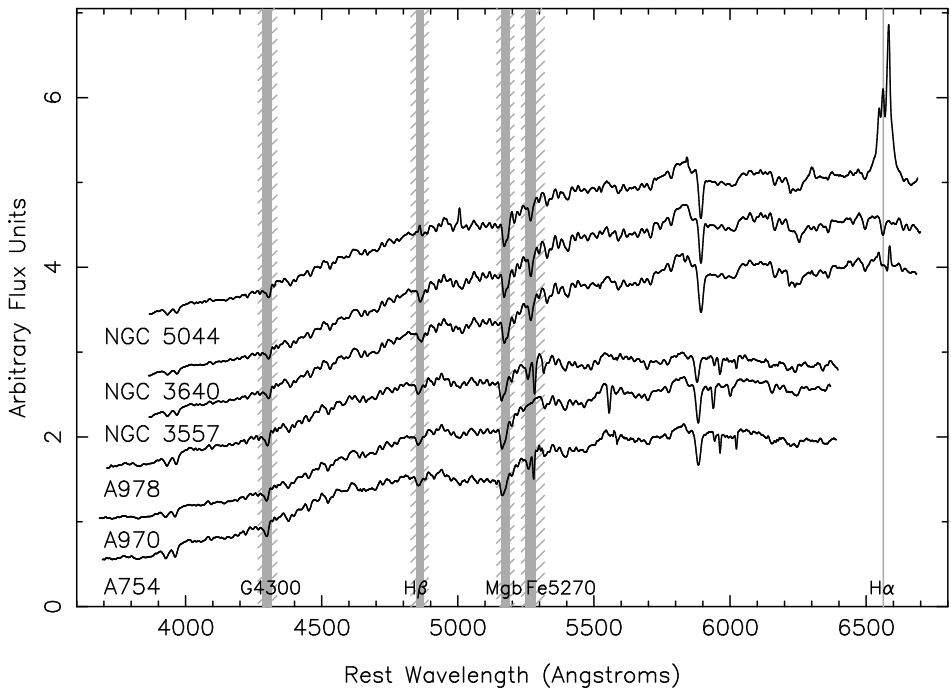}
}  
	}
  \end{center}
\caption{The central spectra for all 6 galaxies, adjusted to the restframe.  
The flux scale is arbitrary. The blue and red continuum (hashed
regions) and central bands (solid regions) of several important Lick
indices are also indicated for information.}
\label{spectra}
\end{figure*}

The spectra of the central aperture of each galaxy are illustrated in
Figure~\ref{spectra}.

\section{Analysis}
\label{section_analysis}

\subsection{Kinematics}

Determination of the recession velocities and velocity dispersions was
carried out by cross-correlating the galaxy spectra with the stellar
spectra, using the IRAF package FXCOR.  As emission lines could affect
the measurement of the kinematics we manually excised emission lines
from the spectra using the IRAF task {\it splot} for this measurement.

We measured velocity dispersions in the central aperture of size
$a_e/8\times$ our slit-width of $0.5^{\prime\prime}$.  These are given
in Table~\ref{kin_table}, along with values from the literature.  The
table also indicates that the literature values are all measured in
apertures of different sizes.  We therefore use the equations provided
in \cite{jorgensen95b} to adjust their apertures to the size used
here: $a_e/8$.  The corrected literature values are also given in
Table~\ref{kin_table}.

\begin{figure}
\begin{center}
     \resizebox{18pc}{!}{
     \rotatebox{-90}{
	\includegraphics{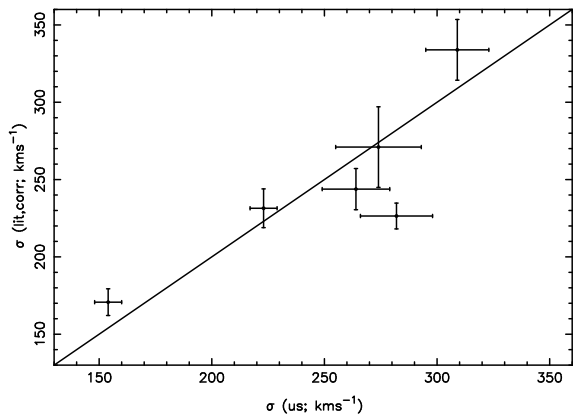}
}  
	}
  \end{center}
\caption{Central velocity dispersion values calculated here compared 
to values from the literature corrected to our aperture size (given in
Table~\ref{kin_table}).  The error bars indicate $1\sigma$ errors. The
solid line indicates the one-to-one relationship.  The absolute mean
difference between our values and those of the literature is
$\mid\sigma$ (us) - $\sigma$ (lit., corr)$\mid=21\pm8$ kms$^{-1}$.}
\label{sigma_comp}
\end{figure}

An accurate comparison of the central value for the velocity
dispersion with literature values is difficult due to the wide range
in systematic errors between studies.  However, we observe an absolute
mean difference between our values and the literature values corrected
to our aperture (Table~\ref{kin_table}) of $21\pm8$ kms$^{-1}$
(Figure~\ref{sigma_comp}).  This is slightly larger than the
statistical errors and provides a measure of the systematic error on
our measurements. The corrected literature values are illustrated
against the observations in
Figures~\ref{n3557},~\ref{n3640},~\ref{n5044},~\ref{a754},~\ref{a970}
and \ref{a978}

\begin{table*}
\begin{center}
\caption{Central kinematics measured within an aperture of
$a_e/8$ and a slit width of $0.5^{\prime\prime}$ and literature
values.}
\begin{tabular}{|llllll|}
\hline
Name&$V_o$&$\sigma_o$&$\sigma_{\rm{lit}}$&Aperture Size&$\sigma_{\rm{lit,corr}}$\\ 
&(kms$^{-1}$)&(kms$^{-1}$)&(kms$^{-1}$)&($^{\prime\prime}$)&(kms$^{-1}$)\\
\hline
NGC 3557 & 3072$\pm$  8&282$\pm$ 16&   244$\pm$ 9 (F)&LS 1$\times$1.5  &226$\pm$  8 \\
NGC 3640 & 1341$\pm$  5&154$\pm$  6&   178$\pm$ 9 (D)&LS 2.5$\times$1.5&171$\pm$  9\\
NGC 5044 & 2755$\pm$ 12&223$\pm$  6&   240$\pm$13 (C)&LS 3$\times$1.5  &231$\pm$ 13\\
A754\#1  &16468$\pm$ 11&309$\pm$ 14&   323$\pm$19 (G)&LS 2$\times$4    &334$\pm$ 20\\
A970\#1  &17512$\pm$  6&264$\pm$ 15&  238$\pm$13 (S) &F 2              &242$\pm$ 13\\
A978\#1  &17695$\pm$ 15&274$\pm$ 19&  260$\pm$25 (O) &LS 8$\times$2    &271$\pm$ 26\\ 
\hline	                             
\label{kin_table}
\end{tabular}
\flushleft The columns indicate (1) galaxy name, (2) central recession velocity, 
(3) central velocity dispersion,(4) Previously measured velocity
dispersion from F \citep{franx89}, D \citep{denicolo05}, C
\citep{carollo93}, G \citep{gorgas90}, S \citep{smith04}, O
\citep{oegerle91}, (5) Aperture type (long-slit, LS, or fibre, F) and
size, (6) literature value, corrected to our aperture size as per
text.
\end{center}
\end{table*}


\subsection{Indices}

Lick indices were measured using the method described in Proctor et
al. (2006; see also
\citealt{proctor04a,proctor04b,proctor05,proctor06}).
\cite{trager98} and \cite{worthey97} index definitions were used throughout.  The indices 
were measured at the wavelength-dependent resolution detailed in
\cite{worthey97}.  

The measured indices require correcting for velocity dispersion
broadening and differences in flux calibration between the Lick system
and these observations.
For galaxy apertures whose velocity dispersions, when combined in
quadrature with the instrumental broadening, resulted in resolutions
higher than that of the relevant index in the Lick system, the spectra
are convolved with an appropriate Gaussian prior to index measurement.
For apertures whose velocity dispersions, when combined in quadrature
with the instrumental broadening, resulted in resolutions lower than
that of the Lick system, indices were corrected using a factor
interpolated from indices measured in the 5 standard stars after
convolving the stellar spectra with a series of Gaussians.  Due to the
varying resolution across the Lick system and the velocity dispersion
profile of galaxies, the second correction is only applied to low
resolution indices ($H{\beta}$ -- Fe5406) measured in the central
apertures (highest velocity dispersions) of our sample.

To correct for the difference in flux calibration between observations
and the Lick system and for the wavelength-dependent resolution of the
Lick system, Lick indices are usually measured for Lick library stars
and compared to the indices measured in the original Lick system
\citep{worthey94}.  As no Lick standards were observed, we cannot 
apply these offsets.  As a compromise, instead, we use the error in
the mean of the offsets measured for 7 Lick standard stars observed by
\cite{proctor04a} to add a measure of the systematic error due to our
lack of flux calibration.
The differences are generally smaller than the scatter and the typical
rms per observation of the Lick calibrators \citep{worthey94}.

The index errors were calculated combining the error from Poisson
noise and the uncertainties in recession velocity and velocity
dispersion.  These are the random statistical errors.  We also add, in
quadrature, the error on the mean of the offset between the
measurements of Lick standard stars and those taken on the Lick/IDS
system from Proctor et al. (2004a; Table~\ref{lick_offset}).  These
provide a measurement of the systematic error due to conversion onto
the Lick system and are of the order of the random errors.

The indices and their errors measured for the central apertures are
given in Appendix~\ref{app_centind} while the indices and errors
measured in the radial apertures are given in
Appendix~\ref{app_apind} for each galaxy.

\begin{table}
\begin{center}
\caption{Offsets from index measurements to those of the Lick system from 
Proctor et al. (2004a; P04a).}
\begin{tabular}{|lcrc|}
\hline
Index&Units&Lick-P04a&Error in Mean\\
\hline
H$\delta_A$ &\AA\ &  0.229 & 0.226 \\ 
H$\delta_F$ &\AA\ & -0.153 & 0.162\\ 
CN$_1$     &mag& -0.014 & 0.013\\ 
CN$_2$     &mag& -0.012 & 0.004\\ 
Ca4227   &\AA\ & -0.089 & 0.060\\ 
G4300    &\AA\ & -0.489 & 0.079\\ 
H$\gamma_A$ &\AA\ & -0.253 & 0.153\\ 
H$\gamma_F$ &\AA\ & -0.118 & 0.064\\ 
Fe4383   &\AA\ & -0.015 & 0.192\\ 
Ca4455   &\AA\ &  0.070 & 0.155\\ 
Fe4531   &\AA\ & -0.382 & 0.091\\ 
C4668    &\AA\ & -0.839 & 0.095\\ 
H$\beta$   &\AA\ & -0.052 & 0.048\\ 
Fe5015   &\AA\ & -0.104 & 0.104\\ 
Mg$_1$     &mag&  0.016 & 0.004 \\ 
Mg$_2$     &mag&  0.023 & 0.003\\ 
Mgb     &\AA\ &  0.024 & 0.043\\ 
Fe5270   &\AA\ & -0.324 & 0.090\\ 
Fe5335   &\AA\ & -0.394 & 0.077\\ 
Fe5406   &\AA\ & -0.181 & 0.045\\ 
Fe5709   &\AA\ & -0.030 & 0.052\\ 
Fe5782   &\AA\ &  0.045 & 0.045\\ 
\hline
\label{lick_offset}
\end{tabular} 
\end{center}
\end{table}

\subsection{Stellar Populations}

We compare the measured Lick indices to predictions from single
stellar population (SSP) models in order to estimate the relative {\it
luminosity-weighted} ages and metallicities of the stellar populations
and also the relative abundances of their $\alpha$-elements.  We use
the luminosity-weighted $\alpha$-element enhancement ratio defined by
\cite{tmb,tmk} - [E/Fe].  This is a proxy for the ratio of the
abundances of the $\alpha$-elements
to the abundance of the iron-peak elements.
Total metallicity is denoted by [Z/H].

The determination of Log (age), [Z/H] and [E/Fe] were carried out by
comparison of the observations to the \cite{tmk} SSP models and
\cite{kmt05} index sensitivities.  The index sensitivities describe the
sensitivity of Lick indices to variations in the chemical abundance
ratios, as computed in stellar atmosphere models.  The SSP values for
each index are interpolated to give a grid of values for $-1<$ Log
(age)$<1.175$ dex and $-1.675<$[Z/H]$<0.5$ dex in 0.025 dex steps.
For each index, at each Log (age), [Z/H] step, values of the index
were also estimated for $-0.3<$[E/Fe]$<0.6$ dex in $\sim0.03$ dex
steps.  The fractional change in each index is calculated assuming
that all elements in the enhanced group vary by the same amount
relative to the Fe elements.  A 3-dimensional grid of model Log (age),
[Z/H] and [E/Fe] values are constructed for each index.  The best fit
Log (age), [Z/H] and [E/Fe] values are then found by simultaneously
fitting the 3-dimensional model grids by $\chi^2$ minimization to as
many observed index values as possible.

Like many fitting routines we employed a $3\sigma$ `clipping' process
so that indices lying $>3\sigma$ from the model fit are clipped and
the fit is re-iterated until there are no indices outside $3\sigma$.
Seeking stability, certain further indices were also clipped from the
fitting, until removal of no single index resulted in changes
significantly different from that implied when including the clipped
index.

The main causes for clipping further indices were:

(1) The GMOS long-slits have two small bridges, of width
$3^{\prime\prime}$, which result in gaps in the spectra - henceforth
termed as `chip gaps'.  Depending on the redshift of each galaxy, we
changed the blaze wavelength of the slit, so as to move these chip
gaps into wavelength ranges without indices. However, this was not
always successful and some indices were lost as a result.

(2) Sky-line residuals lying in regions an index is redshifted into.

(3) Indices whose passbands lie at some distance from the index itself
(e.g. CN$_2$, G4300, Mg$_1$, Mg$_2$) are subject to their continuum
being incorrectly estimated, relative to the index, if the data are not
flux calibrated.

(4) Emission lines contaminate the absorption indices, particularly
the Balmer lines and Fe5015. We chose not
to use simplistic corrections to the absorption indices to resolve
this issue due to their intrinsic uncertainties
\citep{nelan05,kuntschner06}.  Furthermore, the multi-index fitting
method we are using enables us to measure ages from the combined age
dependencies of the remaining indices \citep{proctor04a}.

The extra indices clipped and the reasons for doing so are summarised
in Table~\ref{indice_clip}.

\begin{table*}
\begin{center}
\caption{A summary of indices clipped from the stellar population analysis of each galaxy.  
}
\begin{tabular}{lllll}
\hline
Galaxy&Chip Gap&Sky Line&Flux Calibration&Emission\\
\hline
NGC 3557&C4668, Fe5782&&CN$_2$, G4300&\\
NGC 3640&&&H$\delta_A$, CN$_1$, CN$_2$, G4300&\\
NGC 5044&&&CN$_2$, G4300, Mg$_2$&H$\beta$\\
A754\#1&C4668&Mg$_1$&&\\
A970\#1&&Mg$_1$, Fe5270&G4300&\\
A978\#1&&Mg$_1$&H$\delta_A$, G4300&\\
\hline
\end{tabular} 
\label{indice_clip}
\end{center}
\end{table*}

The errors on the stellar population values were calculated by means
of a Monte Carlo analysis: the model indices associated with the
fitted Log (age), [Z/H] and [E/Fe] value were extracted and a Gaussian
of width equal to the observed index errors was applied to each index.
Index values were then drawn at random from that distribution and the
stellar populations refitted.  This was repeated 50 times and the rms
of the values derived gives the errors we present below.

Previous authors examining GMOS long-slit data have noted that
scattered light in the spectrograph is a significant problem that
would compromise their index measurements \citep{norris06}, while
\cite{proctor05} have shown that scattered light made negligible
difference to their results.  Scattered light would have the effect of
lowering measured index values by applying a constant offset to the
spectra, which would lead to incorrect age gradients at large radii.
It is possible to examine the effects of this problem as the
2-dimensional GMOS spectrum contains 3 unexposed regions: at the
bottom of the slit and 1/3 and 2/3 of the distance along the spatial
direction of the image.  After bias subtraction these regions should
contain no flux, such that any scattered light will be evident here.
\cite{norris06} solve this problem by interpolating across the three
unexposed regions and subtracting that level from the image.  We
fitted a Gaussian to the three unexposed regions in our closest
galaxy, NGC 3640, and subtracted that Gaussian from the 2-dimensional
image.  We then proceeded with our analysis as described above.  We
found that, with the exception of the outermost aperture, the
kinematics and stellar populations measured were within the errors of
our original measurements.  We therefore conclude that scattered light
in the GMOS spectrograph has negligible effect on the results
presented below.

\begin{table}
\begin{center}
\caption{Central values calculated within an aperture of $a_e/8$ and slit-width of $0.5^{\prime\prime}$.}
\begin{tabular}{|lrcr|}
\hline
Galaxy&age$_o$&[Z/H]$_o$&[E/Fe]$_o$\\
&(Gyr)&(dex)&(dex)\\
\hline
NGC 3557   &11.2$\pm$ 0.7& 0.25$\pm$ 0.01&  0.21$\pm$ 0.02\\
NGC 3640   & 7.5$\pm$ 0.6& 0.17$\pm$ 0.02&  0.18$\pm$ 0.02\\
NGC 5044   &15.0$\pm$ 0.9& 0.10$\pm$ 0.02&  0.40$\pm$ 0.03\\
A754\#1    &15.0$\pm$ 0.9& 0.20$\pm$ 0.01&  0.24$\pm$ 0.03\\
A970\#1    &15.0$\pm$ 0.9& 0.22$\pm$ 0.02&  0.27$\pm$ 0.03\\
A978\#1    &11.2$\pm$ 0.7& 0.08$\pm$ 0.04&  0.18$\pm$ 0.04\\
\hline
Mean       &12.1$\pm$1.4&0.17$\pm$0.03&0.25$\pm$0.03\\
\hline
\label{cents_table}
\end{tabular} 
\end{center}
\end{table}

The derived central (within an $a_e/8$ aperture and a slit-width of
$0.5^{\prime\prime}$) velocities and velocity dispersions are given in
Table~\ref{kin_table} and central ages, metallicities and
$\alpha$-index enhancements in Table~\ref{cents_table}.  We note that
the central values and the results we obtain are consistent, within
the errors, for values measured within $r_e/8$.  The observed radial
aperture measurements are given in Appendix~\ref{app_apder} and are
plotted in
Figures~\ref{n3557},~\ref{n3640},~\ref{n5044},~\ref{a754},~\ref{a970}
and \ref{a978}.  The Figures illustrate the data at the position angle
they were observed at.  We have indicated the observed PA and
illustrated the effective radius at those angles (i.e. $a_e$) for each
galaxy.  For those readers interested in the major axis gradients, the
transformation is given in Equation~\ref{rad_eq}.  The gradients
illustrated in these figures were fitted with linear relationships
using a $\chi^2$ minimization with errors in the Y-direction.  To
avoid the uncertainties and radial flattening due to seeing effects,
the data within the central $1^{\prime\prime}$ were excluded from the
fit.

We also note that there are intrinsic uncertainties in the models
themselves that add a systematic error beyond that measured by our
Monte Carlo analysis.  These errors are not relevant when calculating
gradients, or relationships within our sample, however they are
relevant when comparing our results to those of other authors as in
Section~\ref{section_indiv_gals} below and in comparing our
relationships with those of other studies in
Section~\ref{section_central}.  In these sections we therefore add a
factor of $0.1$ dex in quadrature to our Monte Carlo errors as an
indicator of the combined random plus systematic errors.

\section{Individual Galaxies}
\label{section_indiv_gals}

\begin{figure}
\begin{center}
     \resizebox{18pc}{!}{
     \rotatebox{-90}{
	\includegraphics{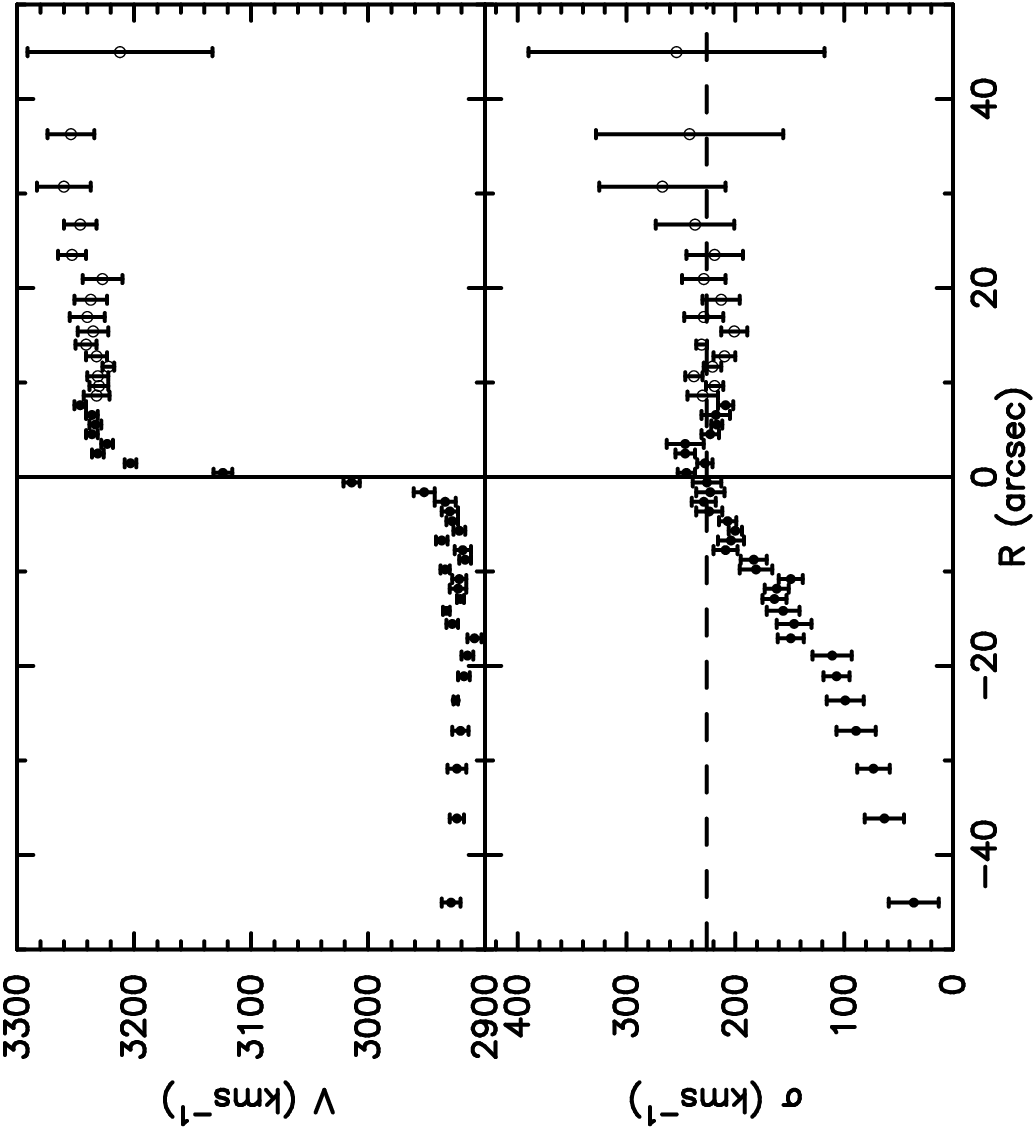}
	\includegraphics{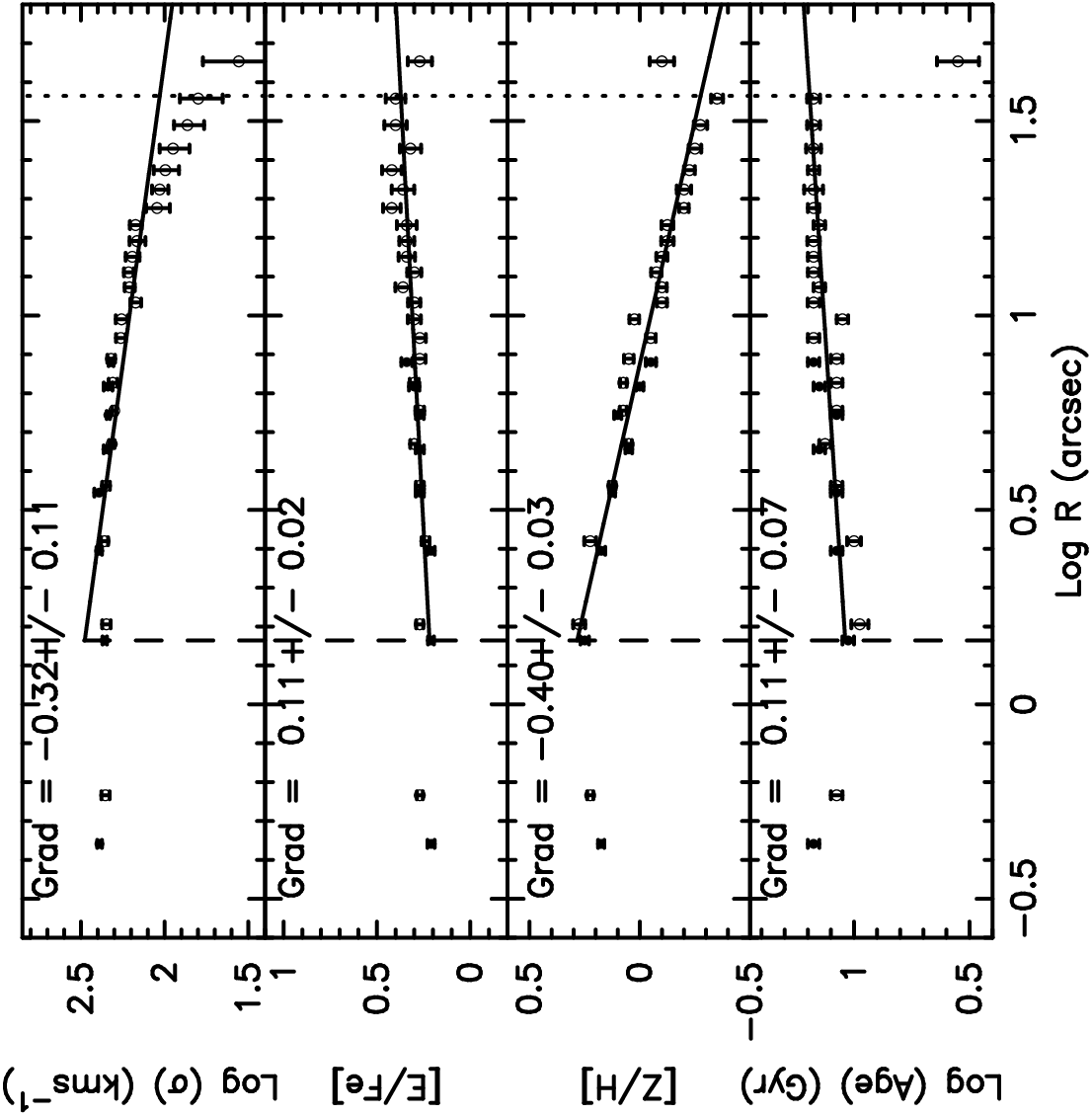}
}  
	}
  \end{center}
\caption{The radial profiles for NGC 3557 measured along the major axis 
of 31$^\circ$ (Table~\ref{obs_table}).  The top two panels show the
recession velocity and velocity dispersion profiles with linear
radius.  The open points in the top two panels indicate those
contaminated by another galaxy in the slit (see text) -- these points
are not used in the gradient calculation.  The errors indicate
$1\sigma$ errors.  The horizontal dashed line in the velocity
dispersion panel indicates the corrected literature value given in
Table~\ref{kin_table}.  The lower 4 panels show measured quantities
with logarithmic radius.  Here the vertical dashed line indicates the
seeing radius, the dotted line indicates the effective radius at the
observed position angle (Table~\ref{obs_table}) and the solid line the
linear fit to the data (whether it is significant or not) with the
gradient and $1\sigma$ error on the gradient given in the top
left-hand corner of each panel.  The open points in the lower 4 panels
indicate the right-hand side of the galaxy and the closed points the
left-hand side. The error bars are the $1\sigma$ errors from the Monte
Carlo analysis described in the text.}
\label{n3557}
\end{figure}

\begin{figure}
\begin{center}
     \resizebox{18pc}{!}{
     \rotatebox{-90}{
	\includegraphics{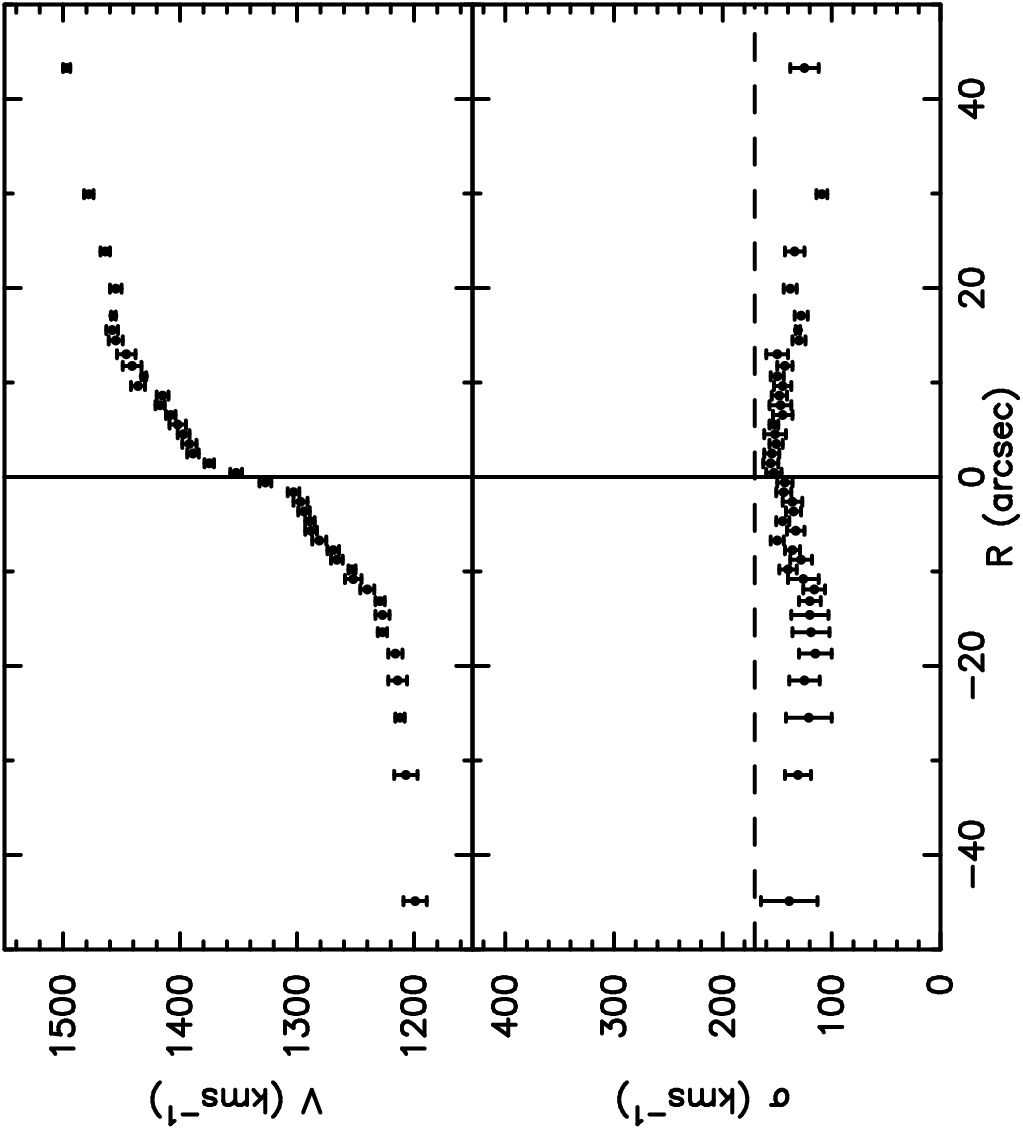}
	\includegraphics{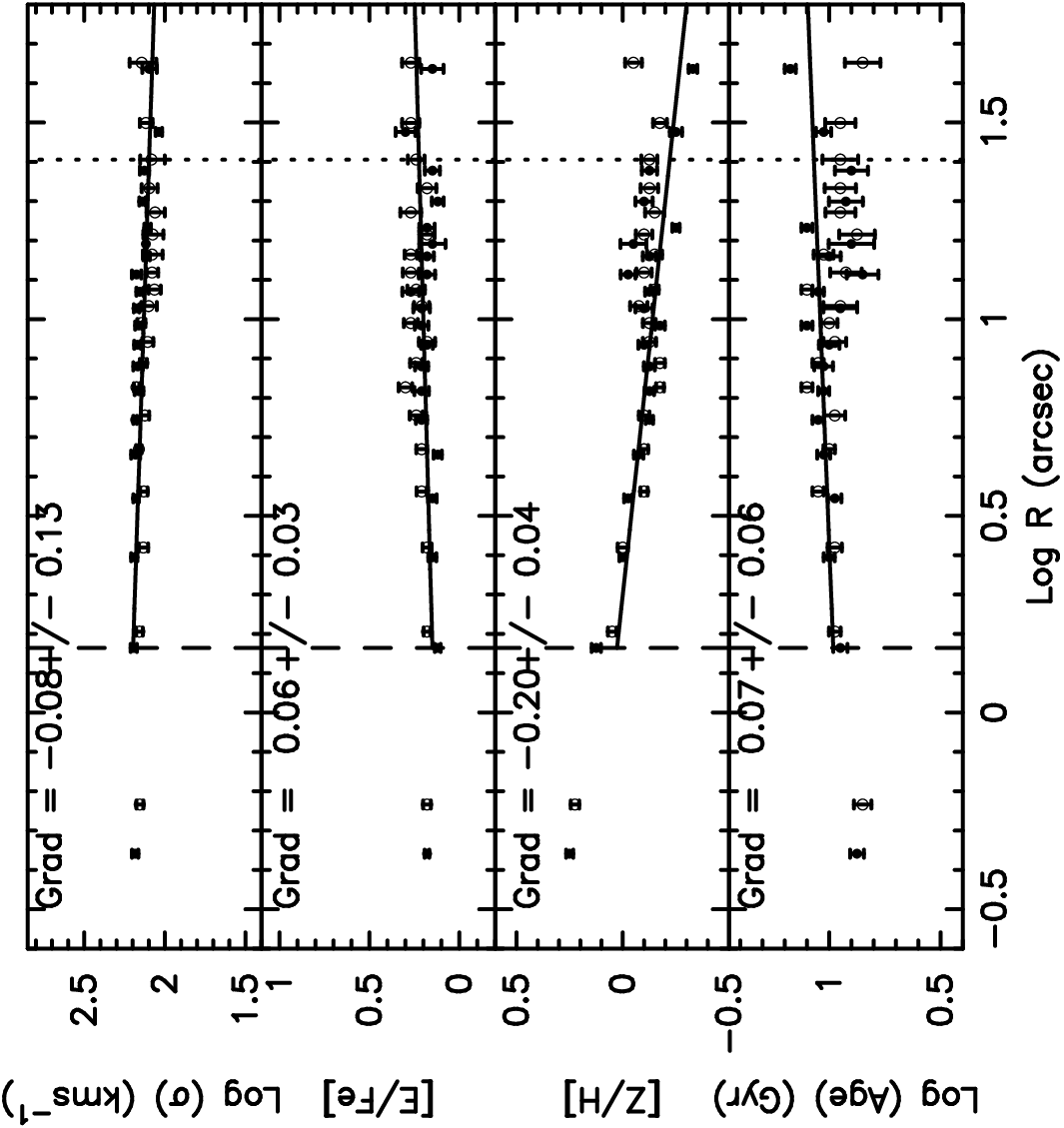}
}  
	}
  \end{center}
\caption{The radial profiles for NGC 3640, measured along the observed axis, 
PA$~=31^\circ$ (major axis$~=95^\circ$; Table~\ref{obs_table}).  The
caption is as for Figure~\ref{n3557}.}
\label{n3640}
\end{figure}

\begin{figure}
\begin{center}
     \resizebox{18pc}{!}{ \rotatebox{-90}{
     \includegraphics{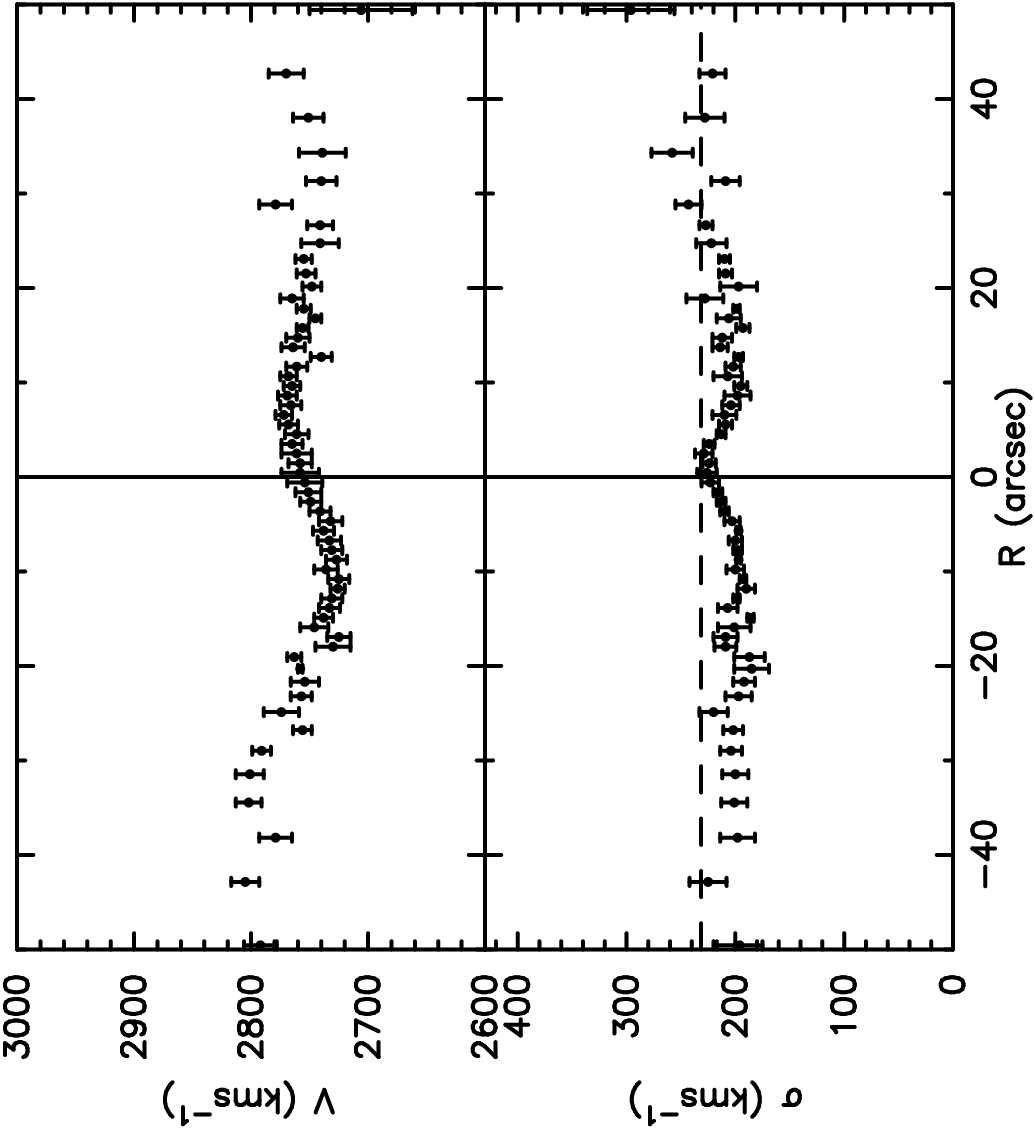}
     \includegraphics{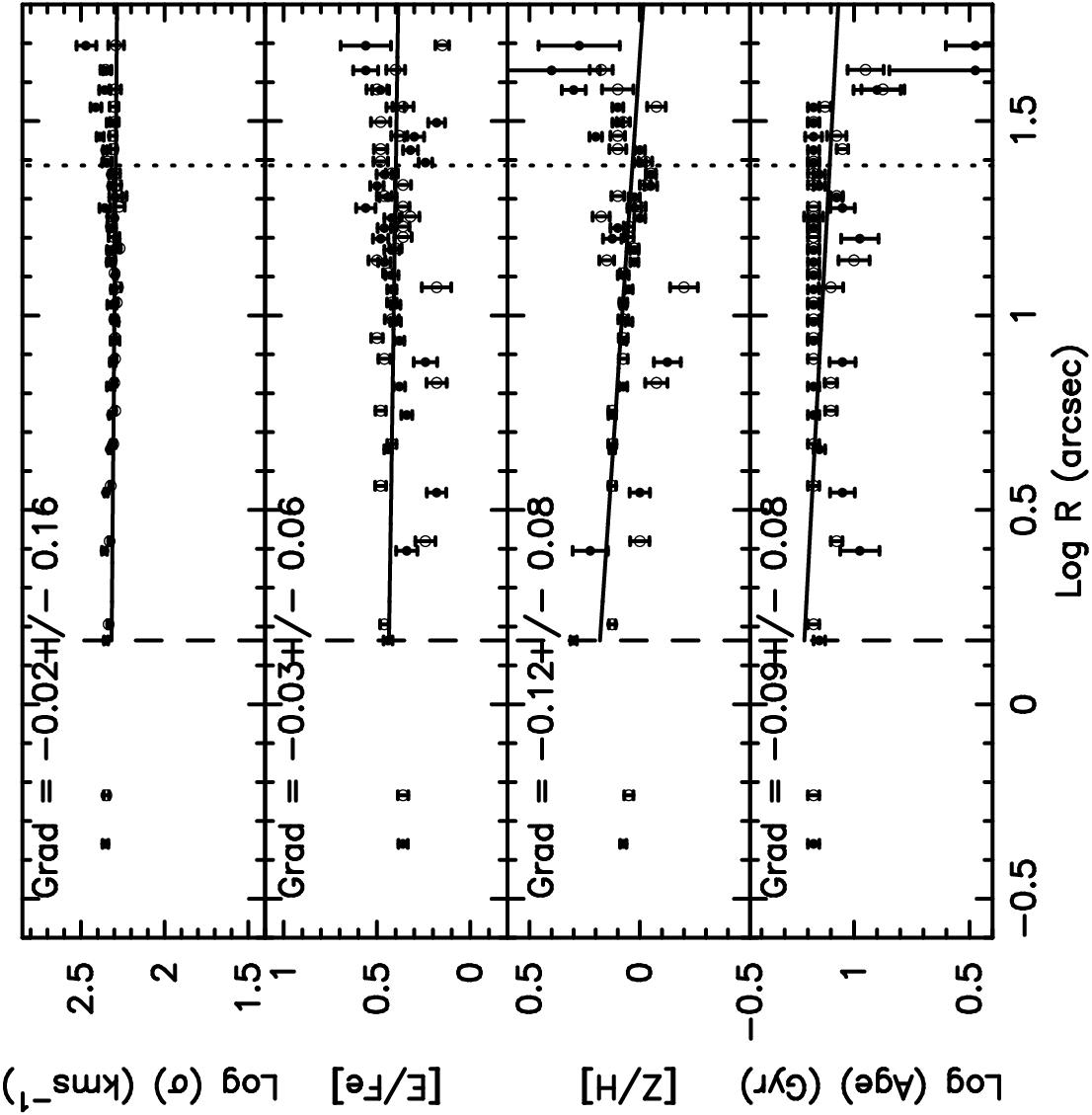}
     } } \end{center}
\caption{The radial profiles for NGC 5044 measured along the major axis 
of 20$^\circ$ (Table~\ref{obs_table}). The caption is as for
Figure~\ref{n3557}.}
\label{n5044}
\end{figure}

The kinematics and stellar populations are plotted in
Figures~\ref{n3557},~\ref{n3640},~\ref{n5044},~\ref{a754},~\ref{a970}
and \ref{a978}.  Here we discuss the results specific to each galaxy.

\subsection{NGC 3557}

NGC 3557 lies at the spatial centre of the NGC 3557 group
\citep{brough06}.  This is a dynamically mature group (e.g. 
\citealt{brough06}) with associated intra-group X-ray emission 
\citep{osmond04}.



The results of the stellar population analysis of this galaxy are
illustrated in Figure~\ref{n3557}.  There is a small galaxy on one
side of our slit which contaminates the spectrum from
$r\geq+8.5^{\prime\prime}$.  These data are illustrated in the top two
panels of Figure~\ref{n3557} as open symbols, and are not included in
the calculation of the gradients.  The stellar populations of NGC 3557
were also studied by \cite{annibali06}: Within an aperture of size
$r_e/8$ and a slit-width of $2^{\prime\prime}$, they found NGC 3557 to
have an age of $5.8\pm0.8$ Gyrs, [Z/H]$=0.034\pm0.004$ dex and
[E/Fe]$=0.17\pm0.02$ dex.  In contrast, within an aperture of $a_e/8$
and a slit-width of $0.5^{\prime\prime}$ we find an age of
$11.2\pm3.5$ Gyrs, [Z/H]$=0.25\pm0.10$ dex and [E/Fe]$=0.21\pm0.10$
dex; i.e. we find an older and more metal-rich centre to this galaxy
but a consistent [E/Fe] ratio.  These differences are not due to the
age-metallicity degeneracy as this would result in us calculating an
older age but a lower metallicity.  \cite{annibali06}
apply an emission correction to their H$\beta$ index
and only use a few indices to measure their ages, metallicities and
$\alpha$-element enhancement ratios.  Given these differences, the
possibility for slit mis-alignment between the two studies, and the
evidence that our method is robust to the effects of emission
\citep{proctor05}, thus avoiding the need to apply uncertain emission 
corrections, we are convinced that our method is the more robust.
Below, we show that when other data are reanalysed using the same
techniques and models used here, the differences between the results
obtained are less significant.


NGC 3557 galaxy shows extreme rotation (Figure~\ref{n3557}).  The
degree by which a galaxy is rotationally supported, in terms of that
expected for a rotationally flattened, oblate spheroid is quantified
by the ratio $(v/\sigma)^{\ast}=(V_{max}/\langle \sigma
\rangle)/[\epsilon/(\epsilon-\epsilon)]^{1/2}$ \citep{davies83}.  This 
is normally $\ll1$ for elliptical galaxies brighter than $M_B>-20$ mag
(i.e. $M_K>-24$ mag), suggesting that these bright galaxies are
supported by velocity anisotropy.  NGC 3557 has $M_K\sim-20.5$ mag and
$(v/\sigma)^{\ast}\sim1$ \citep{fried94}, making it a rotationally
flattened, oblate spheroid.  Numerical simulations suggest that
rotationally supported galaxies are formed by different formation
mechanisms to galaxies that are supported by velocity anisotropy:
Isotropic oblate rotators form either from major (1:1) dissipational
mergers \citep{cox06} or minor (3:1) dissipationless mergers
\citep{naab06}, whilst anisotropic galaxies are formed by major 
dissipationless mergers \citep{cox06,naab06}.  However, NGC 3557 has a
relatively steep metallicity gradient and an older central age,
consistent with forming by collapse at high redshift while the
kinematics may be a result of a dissipationless or dissipative
collapse.

\subsection{NGC 3640}

NGC 3640 is the brightest galaxy in the NGC 3640 group.  This group
appears to be dynamically young as no X-ray emission was detected
above $3\sigma$ of the background level by {\it ROSAT}
\citep{osmond04}.



We found emission in the central arcsecond of this galaxy (consistent
with \citealt{goudfrooij94}), so the H$\beta$ and Fe5015 indices which
are heavily affected by emission were excluded from the analysis of
the central two apertures.  The results are illustrated in
Figure~\ref{n3640}.

\cite{denicolo05} studied this galaxy and found an age 
(within an $r_e/8$ aperture and $1.5^{\prime\prime}$ slit-width) of
$2.5^{+0.7}_{-0.3}$ Gyrs, [Z/H]$=0.53\pm0.13$ dex and [E/Fe]$=0.12$
dex.  In contrast, within an aperture of $a_e/8$ and a slit-width of
$0.5^{\prime\prime}$ our central age $=7.5\pm1.8$ Gyrs, [Z/H]
$=0.17\pm0.10$ dex and [E/Fe] $=0.18\pm0.10$ dex.  We find an older
and less metal-rich centre to this galaxy but a consistent [E/Fe]
ratio, despite using similar models to
\cite{denicolo05} (who use the \citealt{tmb} models with
\citealt{tripicco95} $\alpha$-element enhancements).
\cite{denicolo05} 
apply an emission correction to their H$\beta$ indices and only use a
few indices to measure ages, metallicities and $\alpha$-element
enhancement ratios.  They find their galaxies to be systematically
younger and more metal-rich when comparing to results obtained by
\cite{terlevich02}, even when using the same models.  The 
\cite{denicolo05} data have been reanalysed using the same models 
and techniques as used here, finding an age of $4$ Gyr, [Z/H]$=0.35$
dex and [E/Fe]$=0.06$ dex, within $1.5\sigma$ of our results.  We
therefore conclude that, given the different apertures involved, the
uncertainty of emission corrections and the possibility for slit
mis-alignment between these studies, our results are robust.


The recession velocity profile of NGC 3640 shows indications of a
possible disk in the central $r\sim5^{\prime\prime}$ and of rotation: 
\cite{bender92} found $(v/\sigma)^{\ast}=0.8$ for this galaxy.  This is not 
unusual given that $M_K\sim-24$ mag.  NGC 3640 also has a shallow
metallicity gradient and a younger central age than other galaxies in
the sample.  These results are consistent with NGC 3640 having
undergone a dissipational merger up to 7.5 Gyrs ago.


\subsection{NGC 5044}
\label{n5044_sect2}

NGC 5044 lies at the spatial and kinematic centre of the NGC 5044
group.  This is a large, dynamically-evolved group
(e.g. \citealt{brough06}) with associated intra-group X-ray emission
\citep{osmond04}.  X-ray observations with the {\it XMM} and {\it Chandra} 
satellites by \cite{buote03} and \cite{tamura03}, respectively, have
shown that the hot gas in the NGC 5044 group is dominated by a cooler
component of $T_X\sim0.7$ keV within a radius of 10 kpc
($r\leq63^{\prime\prime}$) and by a warmer component of $T_X\sim1.4$
keV beyond a radius of 40 kpc.  The cooler component is similar to the
kinetic temperature of the stars in the central galaxy suggesting the
presence of stellar ejecta from NGC 5044 itself, while the hot phase
is more characteristic of the halo of the group.

NGC 5044 is particularly interesting as, despite appearing in the
optical as a very ordinary elliptical, it has been found to have very
bright ionized gas emission, extending to radii of $40^{\prime\prime}$
in large filaments \citep{goudfrooij94,machetto96}.  \cite{caon00}
compared the kinematics of the gas emission to that of the stellar
component, finding the gas velocity profile to be very irregular at
the 3 position angles (PA) that they studied and that the gas is
systematically blue-shifted with respect to the stellar velocity
profile by $\sim60-100$ kms$^{-1}$, i.e. the gas is not in dynamical
equilibrium.  They concluded that the gas has recently been accreted
from outside of the galaxy.  



Following \cite{caon00}, we examined the gas emission by fitting a
Gaussian profile to the [NII]$\lambda 6583.6$ line using the IRAF task
{\it splot}.  This line is the strongest and most prominent throughout
the galaxy.  We also examined the H$\alpha$ and [OIII] lines - finding
them to have similar equivalent width and recession velocity profiles
as [NII]$\lambda 6583.6$.

\begin{figure}
\begin{center}
     \resizebox{18pc}{!}
{ \rotatebox{-90}
{
     \includegraphics{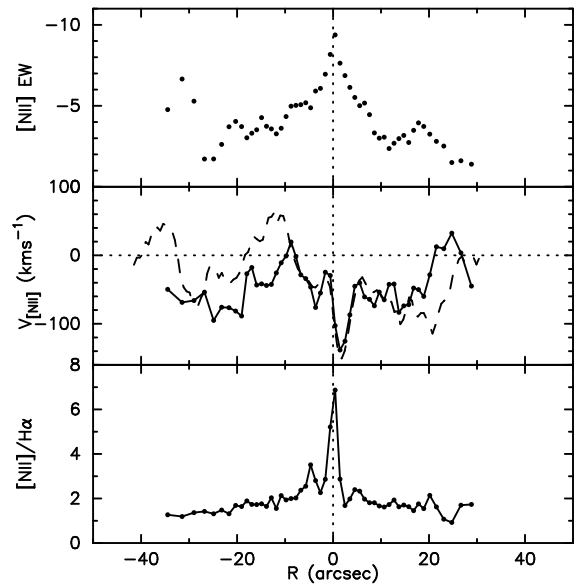}
     } 
} \end{center}
\caption{Distribution of gas emission in NGC 5044 along the major axis 
(PA$~=20^\circ$). The upper panel indicates the distribution of the
equivalent width of the [NII]$\lambda 6583.6$ emission velocity with
radius.  The middle panel indicates the velocity of the [NII] gas
emission, relative to the stellar kinematics.  The solid line
indicates our data, while the dashed line gives the information from
Caon et al. (2000), measured along an axis of PA$~=10^\circ$.  The
lower panel indicates the ratio of [NII] to H$\alpha$ emission.}
\label{n5044_gasprofile}
\end{figure}

Figure~\ref{n5044_gasprofile} shows the equivalent width of the
[NII]$\lambda 6583.6$ line and the gas kinematics we derived, with
respect to those from Caon's PA $=10^{\circ}$ spectra.  We observe
significant emission across all radii we study, including highly
broadened H$\alpha$ emission in the central two apertures
(Figure~\ref{spectra}).  We also observe the same central
blue-shifting of the emission as \cite{caon00}, despite the different
position angle observed.  The blue-shifted gas emission with respect to the
stellar suggests that there is an outflow of material over these
radii.  This could be from a jet linked to the low-luminosity (flux at
1.4 GHz $=0.0347$ Jy) compact ($r\sim1^{\prime\prime}$) nuclear radio
source observed by \cite{condon98} and the X-ray observations of
\cite{buote03}.  The offset in emission velocities between this study
and that of \cite{caon00} outside the central $5^{\prime\prime}$ is
easily understood as a result of the different position angles
observed and the filamentary nature of the observed H$\alpha$
emission.

In order to determine whether the emission from NGC 5044 is due to
star formation or an AGN we 
examine the radial distribution of the [NII]/$H\alpha$ ratio
(Figure~\ref{n5044_gasprofile}) and find it to be $\geq1.5$ over most
of the galaxy, peaking at 3-6 in the central $r<5^{\prime\prime}$.
Examining the diagnostic diagrams of \cite{kewley06} it is clear that
all galaxies with [NII]/$H\alpha>1$ are AGN.  Our findings are
consistent with \cite{rickes04} who concluded that in the central
$r<1$ kpc the ionization source must be an AGN.

The presence of a low-luminosity nuclear radio source and
low-temperature central X-ray component, in addition to the broadened
H$\alpha$ emission, blue-shifted gas emission and high [NII]/$H\alpha$
ratio we observe here lead us to believe that NGC 5044 contains a
low-luminosity AGN at its centre, with a jet pointing towards us.


Due to the presence of emission throughout this galaxy, and the
resulting exclusion of the Balmer lines, this was one of the more
difficult galaxies to fit, resulting in a noisy profile
(Figure~\ref{n5044}).
We do, however, note good agreement with \cite{rickes04} and
\cite{annibali06} for the old central age of this galaxy:
\cite{rickes04} fit 3 old components (age$~=10$ Gyrs) with different
metallicities to the spectrum of NGC 5044 and compare that fit to one
with solar metallicity and 3 different age components, finding that
the spectra are better fitted by a range in metallicity than a range
in age.  \cite{annibali06} use a similar comparison of Lick indices to
SSP models as we do here and, within an aperture of $r_e/8$ and a
$2^{\prime\prime}$ slit-width, they find NGC 5044 to have an age of
$14.2\pm10$ Gyrs, consistent with our central age of $=15.0\pm3.9$
Gyrs (this is the maximum age modelled).  Their $\alpha$-element
enhancement ratios and metallicities are also consistent with ours:
[E/Fe]$_{A06}=0.34\pm0.17$ dex; [E/Fe]$_{us}=0.40\pm0.10$ dex and
[Z/H]$_{A06}=0.015\pm0.022$ dex; [Z/H]$_{us}=0.10\pm0.10$ dex.

\cite{caon00} observe counter rotation of the central regions of 
NGC 5044 with respect to the outer regions: 
The inner $r\leq15^{\prime\prime}$ is counter-rotating in their
PA$=10^{\circ}$ spectra.  This is consistent with the
$r\sim10^{\prime\prime}$ region counter-rotating in our spectra
(PA$=20^{\circ}$).  They also see signs of a further nuclear
counter-rotating component within this region of radius
$\sim2^{\prime\prime}$.  We do not observe this nuclear
counter-rotating component.  However, we do note that the errors in
our recession velocity and velocity dispersion measurements increase
significantly within this region, suggesting that we do not have the
velocity resolution to resolve this structure.

Despite the counter-rotating centre and gas emission, we found no
correlation of age, metallicity or $\alpha$-element enhancement with
these features.  The lack of a visible population of young stars,
despite significant emission, is further evidence for this being an
AGN and not a starburst galaxy.  NGC 5044 also has a very shallow
metallicity gradient, suggesting it has undergone at least one, likely
to be dissipative, merger event.


\begin{figure}
\begin{center}
     \resizebox{18pc}{!}{
     \rotatebox{-90}{
	\includegraphics{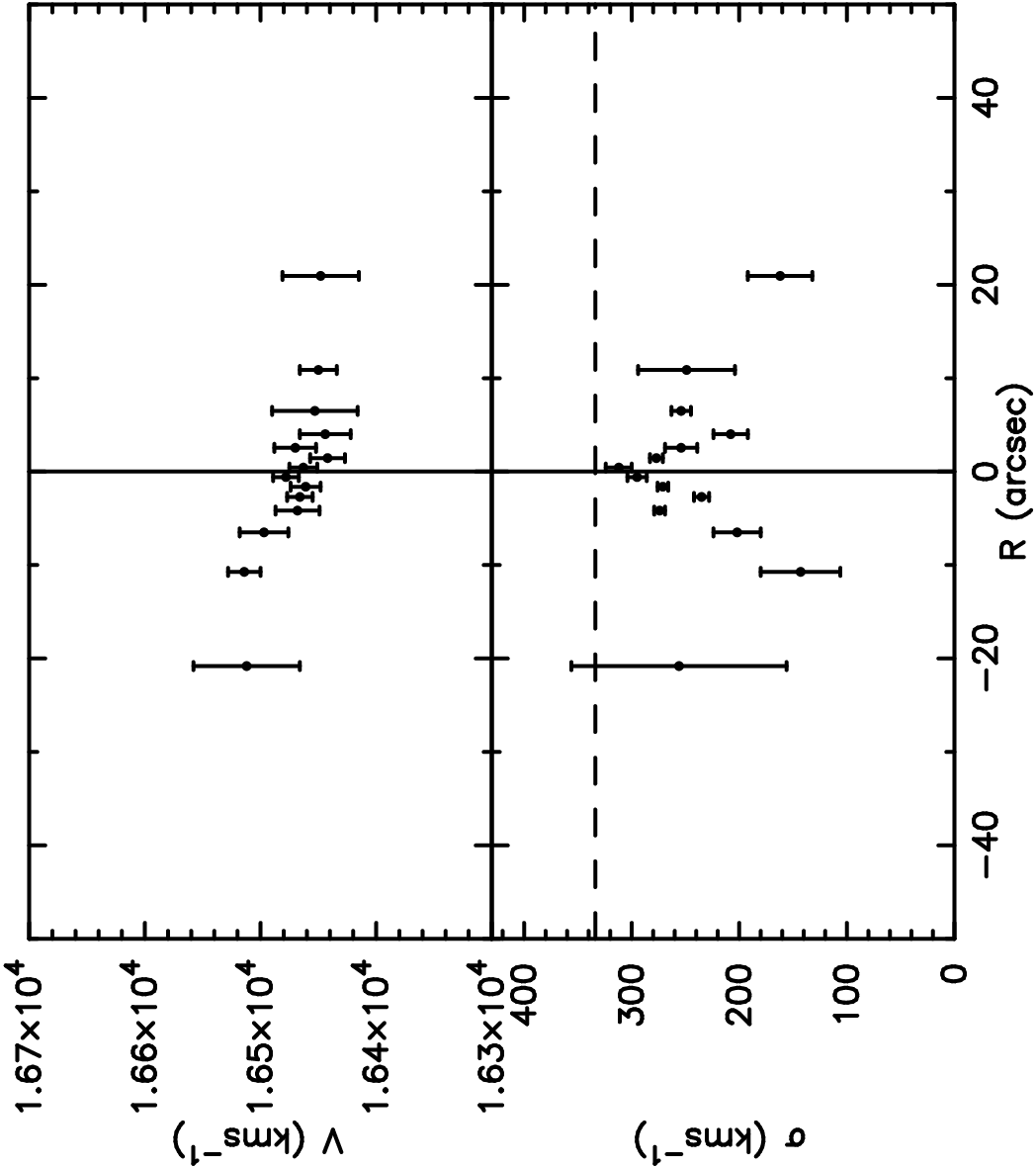}
	\includegraphics{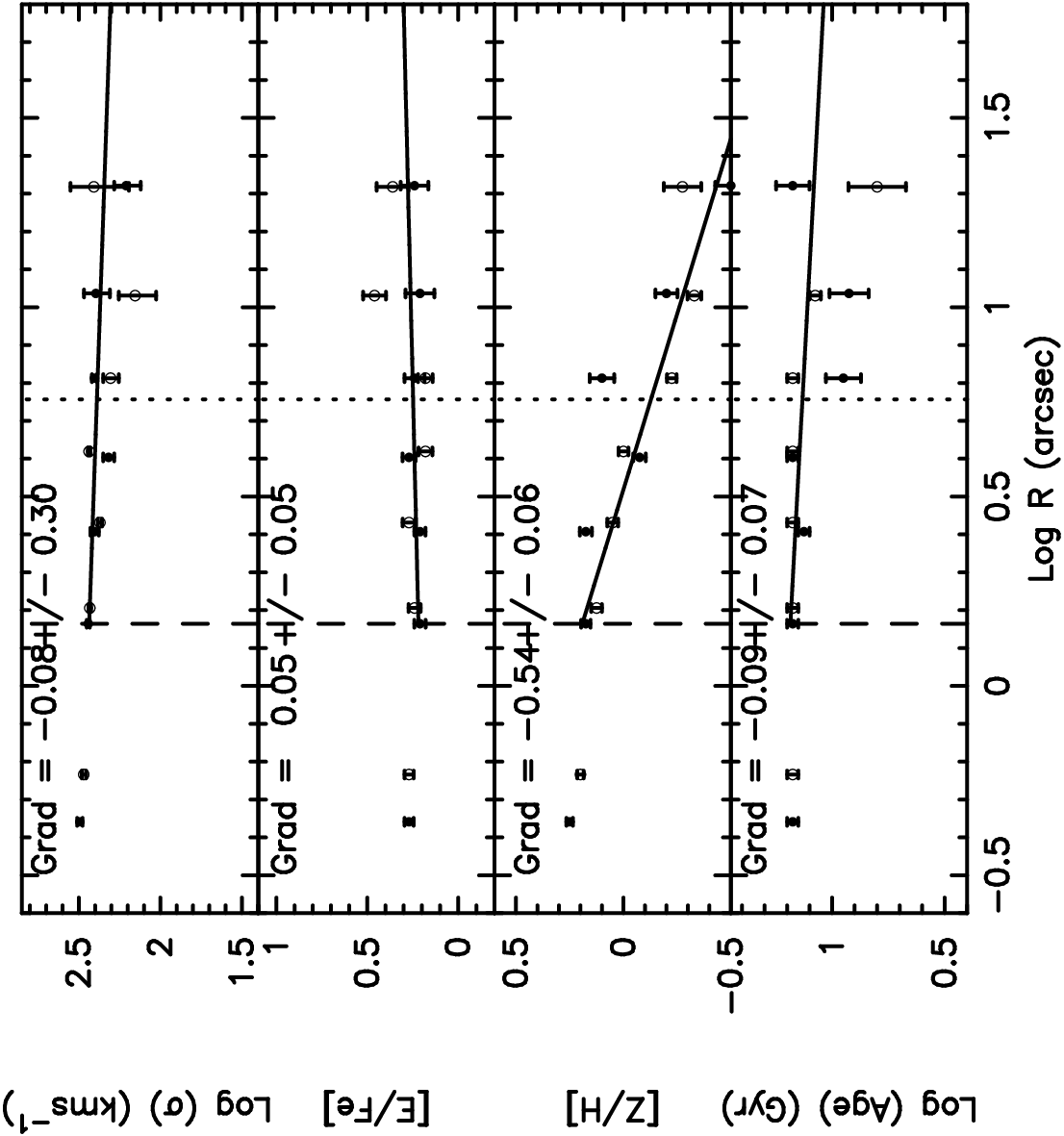}
}  
	}
  \end{center}
\caption{The radial profiles for A754\#1, measured along a PA of $96^\circ$ 
(major axis$~=125^\circ$; Table~\ref{obs_table}).  The caption is as
for Figure~\ref{n3557}.}
\label{a754}
\end{figure}

\begin{figure}
\begin{center}
     \resizebox{18pc}{!}{ \rotatebox{-90}{
     \includegraphics{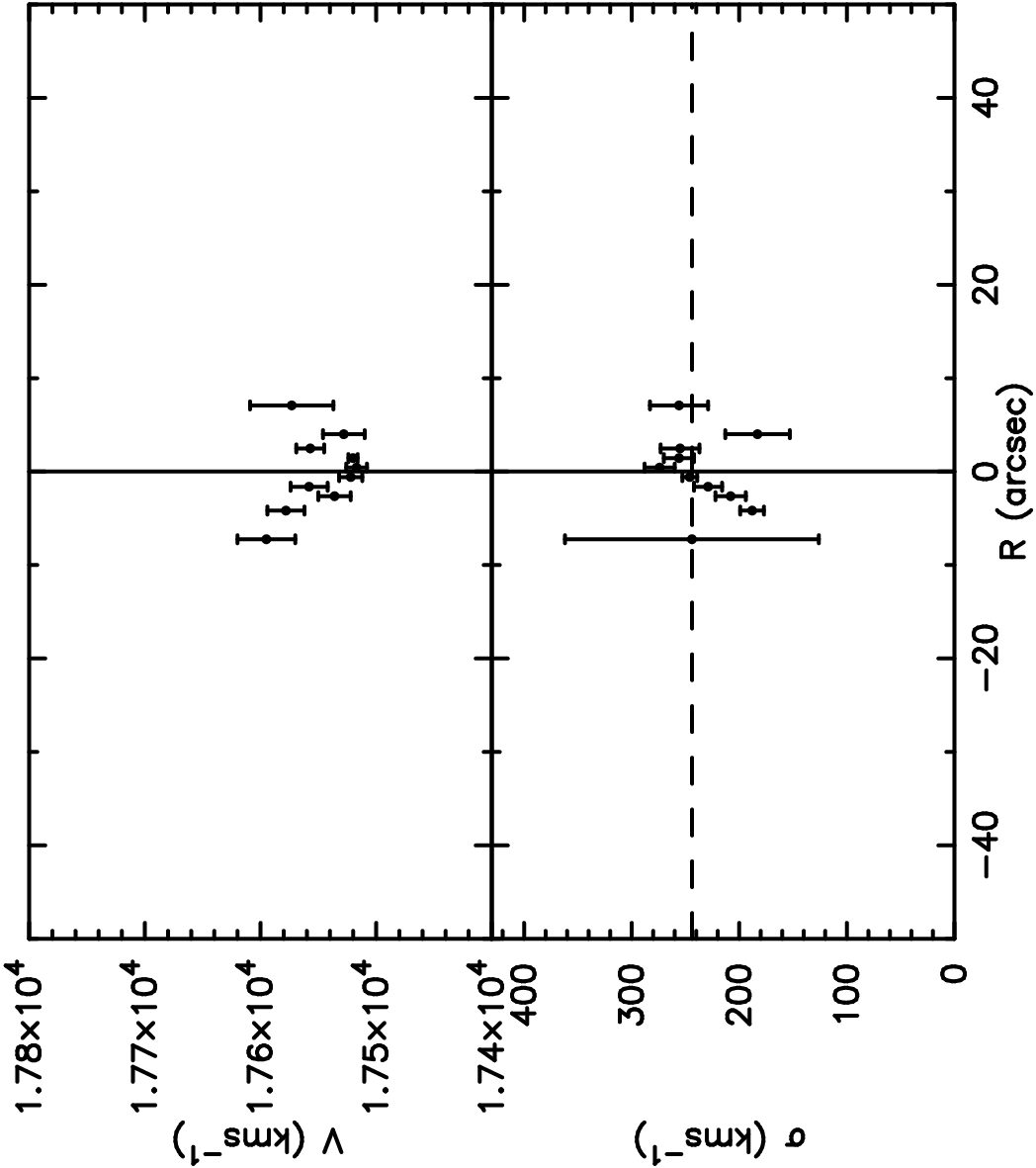}
     \includegraphics{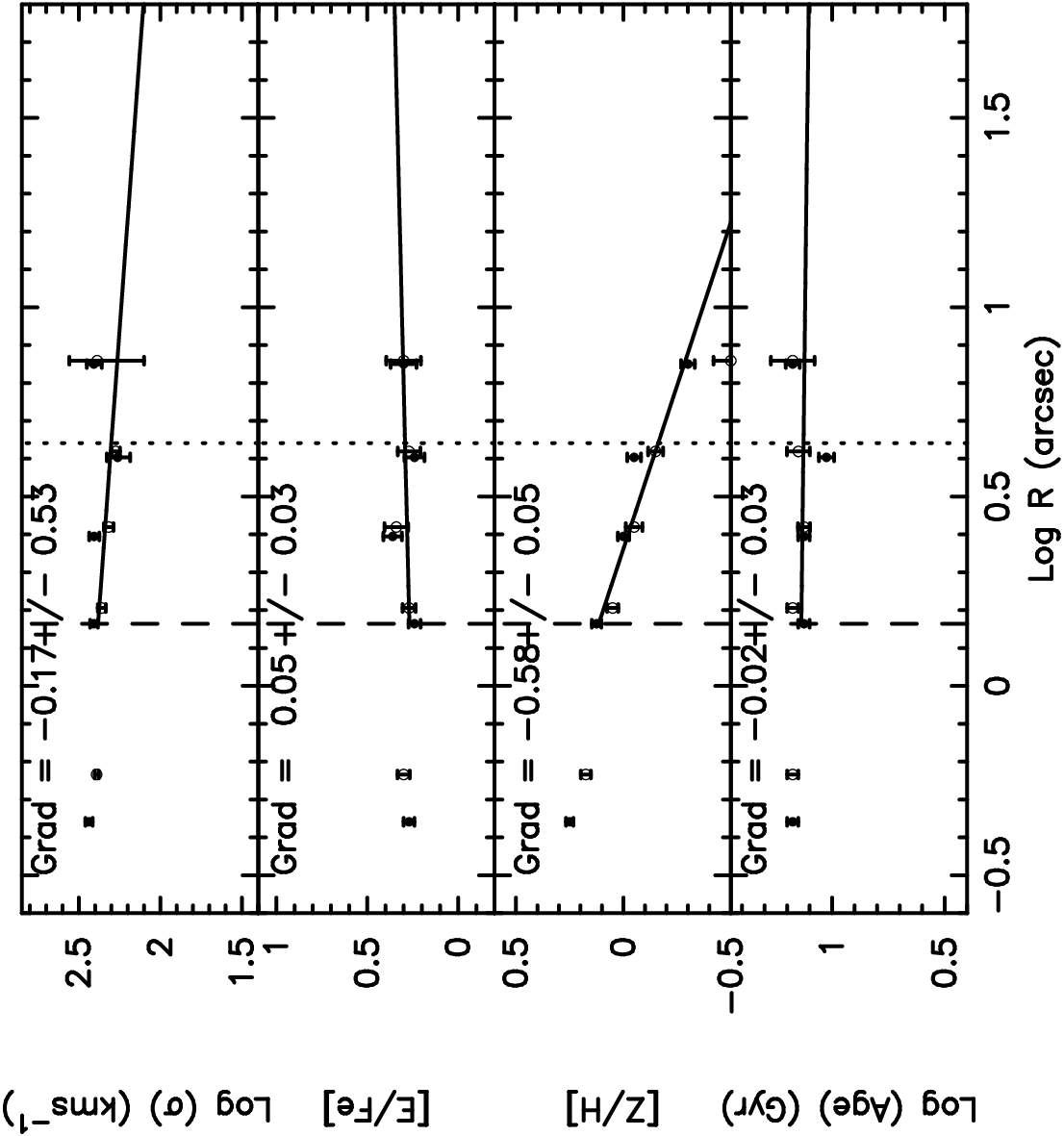} } } \end{center}
\caption{The radial profiles for A970\#1, measured along a PA of $177^\circ$ 
(major axis$~=45^\circ$; Table~\ref{obs_table}).  The caption is as for
Figure~\ref{n3557}.}
\label{a970}
\end{figure}

\begin{figure}
\begin{center}
     \resizebox{18pc}{!}{
     \rotatebox{-90}{
	\includegraphics{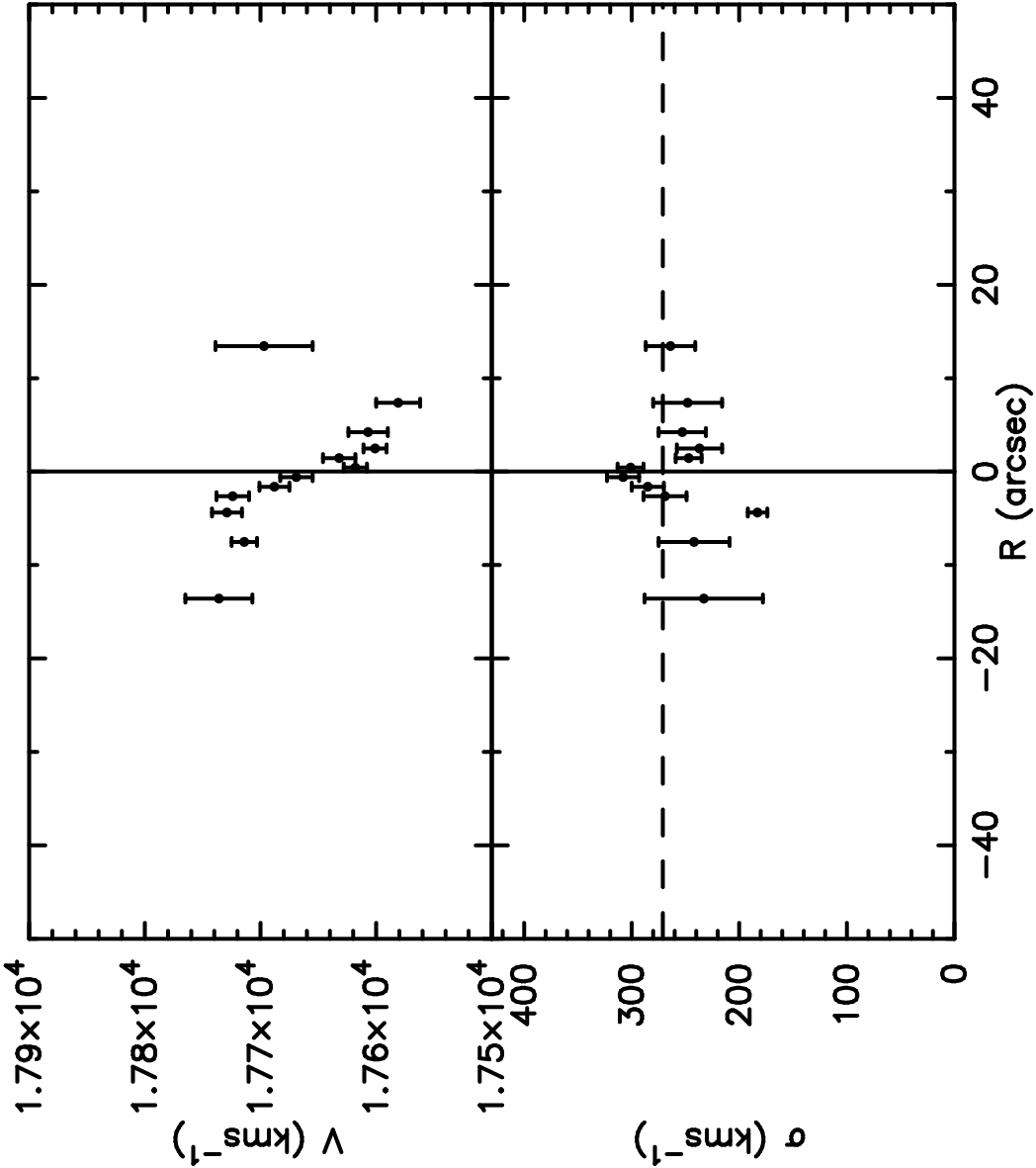}
	\includegraphics{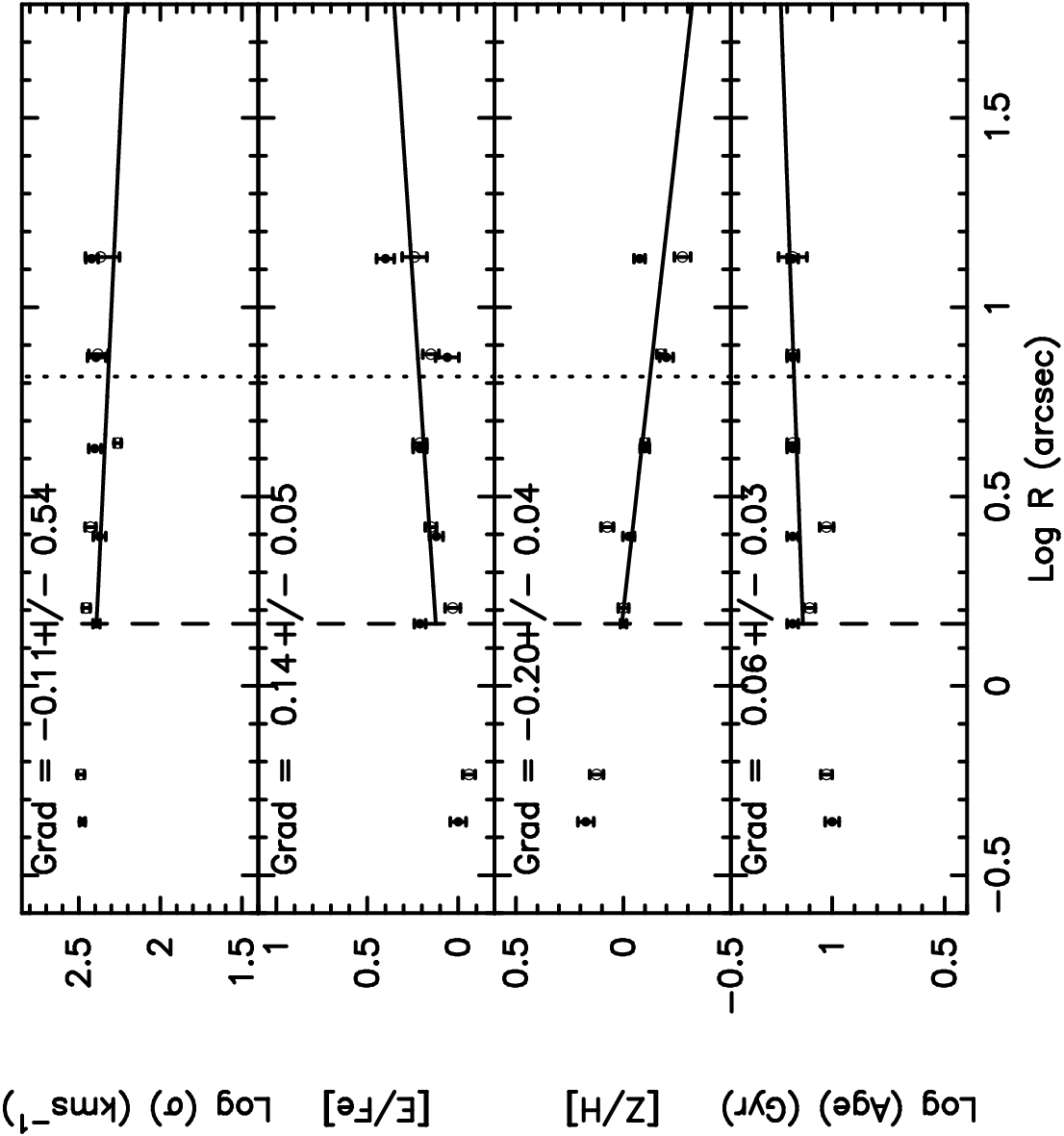}
}  
	}
  \end{center}
\caption{The radial profiles for A978\#1, measured along a PA of $18^\circ$ 
(major axis$~=10^\circ$; Table~\ref{obs_table}).  The caption is as
for Figure~\ref{n3557}.}
\label{a978}
\end{figure}

\subsection{A754\#1}

The Abell 754 cluster consists of two major condensations - the most
massive to the North-West containing the BCG and another $16^{\prime}$
to the South-East, consistent with there having been a recent
collision of two smaller clusters \citep{zabludoff95}.  It also
contains a weak cooling flow \citep{white97}.




No previous age or metallicity measurements have been made of this
galaxy or the other two BCGs of this sample.

We find this galaxy to be uniformly old (Figure~\ref{a754}), with a
steep metallicity gradient, suggesting that the BCG has not undergone
any merger event since its formation at high redshift.  This also
suggests that the BCG has not been affected by the cluster-cluster
merger observed in the distribution of the X-ray emitting gas and
optically observed galaxies by \cite{zabludoff95}.

\subsection{A970\#1}

The Abell 970 cluster, together with Abell 978, is a member of the
Sextans supercluster \citep{einasto97} and was found to have a weak
cooling flow by \cite{white97}. The cluster was studied in detail by
\cite{sodre01} who found A970\#1 to lie at the spatial and kinematic 
centre of the cluster.  They also found evidence for a galaxy group
having recently fallen into the cluster.






We find A970\#1 to be uniformly old with a steep metallicity gradient
(Figure~\ref{a970}).  This suggests that, like A754\#1, A970\#1 has
not undergone any merger event since its formation at high redshift.
There are no signs of star formation, suggesting that the infall of
the group into this cluster has had no effect on the BCG.

\subsection{A978\#1}

A978\#1 was found to have a recession velocity of $V_0=16263\pm44$
kms$^{-1}$ by the ESO Nearby Abell Cluster Survey \citep{katgert96}.
This places it at the kinematic centre of the Abell 978 cluster as
defined by \cite{struble99}.

The Abell 978 cluster has been found to have a cooling flow
\citep{sarazin86} but no cold molecular gas was observed to be
associated with it \citep{salome03}, suggesting that the cooling flow
is not cooling sufficiently to form stars.





A978\#1 shows strong rotation (Figure~\ref{a978}), very unusual in
such a massive galaxy ($M_K\sim-26.5$ mag).  We also find it to have a
slightly younger central age, consistent with the observed low central
$\alpha$-element enhancement ratio.  We also note its shallow
metallicity gradient.  These results all suggest that this galaxy has
undergone a merger event.  If this merger event was dissipative it
must have happened $leq10$ Gyrs ago, however if it was a
disspationless merger it could have occurred more recently.



\section{Relationship with Environment}
\label{section_env}

\begin{figure}
\begin{center}
     \resizebox{21pc}{!}{ \rotatebox{-90}{
     \includegraphics{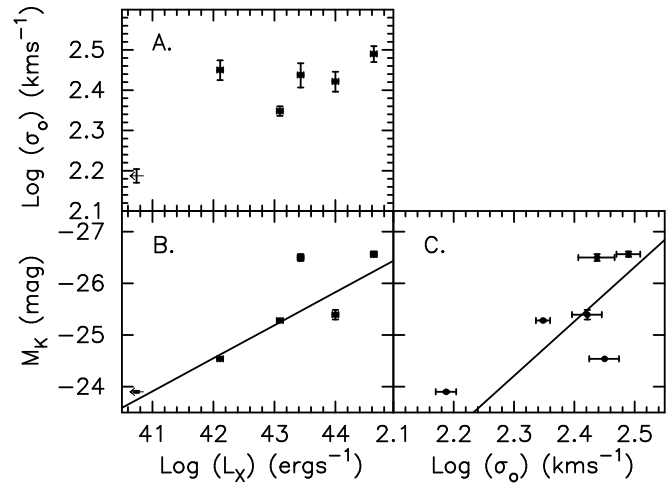}
     } } \end{center}
\caption{The relationship between the X-ray luminosity of the host 
system, $L_X$, and proxies for galaxy mass: central velocity
dispersion ($\sigma_0$, panel A) and absolute $K$-band magnitude
($M_K$, panel B) and the relationship between the two mass proxies,
i.e. the Faber-Jackson relation (panel C).  The error bars indicate
$1\sigma$ errors, the arrow indicates the upper limit in $L_X$
measured for the NGC 3640 group.  The solid lines indicate linear fits
to our data that are statistically significant (all the fits are given
in Table~\ref{env_fits_table}). 
}
\label{global_env}
\end{figure}

\begin{table}
\begin{center}
\caption{Linear fits to the relationships with environment.  The significant relationships are illustrated in Figure~\ref{global_env}.}
\begin{tabular}{|lrll|}
\hline
Relationship (Y - X)&Gradient&Intercept&rms\\
\hline
Log($\sigma_{0}$) - Log ($L_X$)&   0.07$\pm$ 0.03& -0.65$\pm$ 0.08& 0.08\\
$M_K$ - Log ($L_X$)&  -0.64$\pm$ 0.17&  2.32$\pm$ 4.40& 0.59\\
$M_K$ - Log($\sigma_{0}$)&  -7.13$\pm$ 0.07& -8.33$\pm$ 1.98& 0.80\\

\hline
\label{env_fits_table}
\end{tabular} 
\end{center}
\end{table}


Here we examine the dependence of BGG and BCG mass on the X-ray
luminosity of their host group/cluster environment.  X-ray luminosity
is directly related to the density of the intra-group/cluster medium
and therefore provides a quantitative measure of the density of the
environment these galaxies reside in.  In Figure~\ref{global_env} we
compare the central velocity dispersions
and absolute $K$-band magnitudes of the galaxies with the X-ray
luminosity of their host environment.  We fit linear relationships to
these data using $\chi^2$-minimization with errors in the Y-direction.
We note that we only have an upper limit for the X-ray luminosity of
the NGC 3640 group.  It is possible to use the statistical tool of
survival tasks to take upper (and lower) limits into account when
fitting data (e.g. \citealt{osullivan01,brough06}).  However, these
tasks rely on there being values below the upper limit given and also
assume that the upper limits are randomly distributed with respect to
the independent parameter, neither of which are true here.  We
therefore fit to the X-ray luminosity of NGC 3640 as we do to the
other X-ray luminosities, rather than as an upper limit.

The fits are given in Table~\ref{env_fits_table}.  
As a measure of whether or not these relationships are significantly
different from the null hypothesis of no relationship, we take a
relationship to be significant if the fitted gradient is at least a
factor of 3 greater than the $1\sigma$ error on that gradient
(i.e. $\geq3\sigma_{gradient}$).  In this and other plots in this
paper, we only plot the significant
relationships.  We also examined the relationships with the dynamical
mass of these galaxies, measured as $\propto \sigma_0^2r_e$, and did
not find these relationships to be more or less significant than the
relationships with $\sigma_0$.  We therefore only illustrate the
relationships with $\sigma_0$ in this paper.  The individual panels
are discussed separately below:

{\bf A. Log ($\sigma_0$) -- Log $L_X$:} 
We see the suggestion of a trend between the central velocity
dispersions of these galaxies and the X-ray luminosity of their host
cluster: Galaxies in higher density clusters are more massive.  The
BCGs generally have higher velocity dispersions than the BGGs
(Table~\ref{kin_table}).  However, this relationship is not
statistically significant (i.e. only $2\sigma_{gradient}$).
  


{\bf B. $M_K$ -- Log $L_X$:} We observe that more X-ray luminous
systems contain brightest group/cluster galaxies that are
intrinsically brighter in the $K$-band than less X-ray luminous
systems.  This is consistent with studies of BCGs by \cite{edge91} and
\cite{fisher95a}, and the results for BGGs found by \cite{brough06}.
However, it is in contrast to \cite{brough02} who found BCGs at
redshifts $<0.1$ in clusters with X-ray luminosities
$L_X<1.9\times10^{44}$ erg s$^{-1}$ to exhibit a wide range of scatter
in their absolute $K$-band magnitudes.  The BCGs studied here are part
of the \cite{brough02} sample, and therefore lie within the scatter they
observed.  However, the \cite{brough02} sample was significantly
larger (92 BCGs with $z<0.1$).
In \cite{brough06} the correlation of the $K$-band luminosities of
BGGs with the X-ray luminosity of their host group was driven by
groups without extended group-scale X-ray emission detected (i.e. the
X-ray luminosity of their host group was either solely that associated
with the BGG itself or X-rays were not detected above
$3\sigma_{background}$).  We therefore conclude that the difference
between this study and \cite{brough06}, and that of \cite{brough02} is
due to the smaller sample studied here and the inclusion of a BGG in a
group without extended group-scale X-ray emission.  We intend to
investigate this issue in more detail with a significantly larger
sample in a future paper.




{\bf C. Log ($M_K$ -- $\sigma_0$):} Given that we are assuming that
both central velocity dispersion and absolute $K$-band magnitude are
good proxies for galaxy mass, we examine the relationship between them
(the Faber-Jackson relation; \citealt{faber76}) here.  BCGs have
previously been found to lie off the Faber-Jackson relation for normal
ellipticals at optical wavelengths, having brighter magnitudes than
would be predicted by their velocity dispersions
(e.g. \citealt{oegerle91,bernardi06b}).  
Figure~\ref{global_env} illustrates that we find galaxies with larger
velocity dispersions to be brighter, such that
$M_K\propto\sigma^{-10.5\pm0.06}$.  

Globally the only differences between BGGs and BCGs are the lower
velocity dispersions and fainter absolute $K$-band magnitudes and,
therefore, lower mass of the BGGs.

From the predictions of dLB we would expect BCGs to increase in mass
in step with their host system.  The trend of absolute magnitudes with
cluster X-ray luminosity is, therefore, consistent with their
predictions.  However, it is surprising to note that galaxy velocity
dispersion is only weakly related to cluster X-ray luminosity, as we
would expect a much stronger correlation than this from the
semi-analytic predictions.

\section{Central Stellar Populations}
\label{section_central}

Dissipative collapse scenarios predict that elliptical galaxies formed
as a result of the collapse of gas clouds at $z\geq3$ and have evolved
quiescently since
(e.g. \citealt{eggen62,larson74,carlberg84,arimoto87}).  In this
scenario we expect to observe consistently old ages, with
metallicities that scale with mass, high $\alpha$-enhancement ratios
and no dependence of these quantities on environment.  In the
hierarchical model of structure formation, elliptical galaxies are
formed by mergers and accretion events over a Hubble time. If there is
sufficient gas present in the progenitor galaxies then the mergers
will induce star formation.  dLB predict dissipationless mergers in
the recent assembly of BCGs. Therefore, if their predictions are
correct, we would expect to observe no correlation of metallicities
with environment and no signs of star formation in the last $\sim5$
Gyrs.

In the sections below we present the stellar population trends
exhibited by our sample of galaxies.  To compare with studies of the
central values of ages and metallicities in normal early-type galaxies
we use the measurements derived for the inner $a_e/8$ of our galaxies
(i.e. an aperture $a_e/8\times0.5^{\prime\prime}$).
These values are given in Tables~\ref{kin_table}
and~\ref{cents_table}.  We note that three of our galaxies have hit
the maximum age modelled of 15 Gyrs (Log (age)$=1.175$ dex).


\begin{table}
\begin{center}
\caption{Linear fits to the relationships between the central values. 
The significant relationships from the upper section of this table are
illustrated in Figure~\ref{env_central}.  The significant
relationships from the lower section of this table are illustrated in
Figure~\ref{central_values}.}
\begin{tabular}{|lrll|}
\hline
Relationship (Y-X)&Gradient&Intercept&rms\\
\hline
Log(age$_{0}$) - Log ($L_X$)&   0.07$\pm$ 0.01& -1.94$\pm$ 0.04& 0.06\\
$[$Z/H$]$ - Log ($L_X$)&   0.00$\pm$ 0.00&  0.16$\pm$ 0.02& 0.08\\
$[$E/Fe$]$ - Log ($L_X$)&   0.02$\pm$ 0.01& -0.62$\pm$ 0.01& 0.09\\
\hline
Log(age)$_{0}$ - Log($\sigma_{0}$)&   0.78$\pm$ 0.05& -0.78$\pm$ 0.05& 0.09\\
$[$Z/H$]_{0}$ - Log($\sigma_{0}$)&   0.13$\pm$ 0.04& -0.14$\pm$ 0.02& 0.08\\
$[$E/Fe$]_{0}$ - Log($\sigma_{0}$)&   0.04$\pm$ 0.05&  0.15$\pm$ 0.01& 0.09\\
$[$Z/H$]_{0}$ - Log(age$_{0}$)&   0.01$\pm$ 0.04&  0.16$\pm$ 0.03& 0.08\\
$[$E/Fe$]_{0}$ - Log(age$_{0}$)&   0.45$\pm$ 0.03& -0.24$\pm$ 0.01& 0.07\\

\hline
\label{cents_fits_table}
\end{tabular} 
\end{center}
\end{table}

\begin{figure}
\begin{center}
     \resizebox{16pc}{!}{ \rotatebox{-90}{
     \includegraphics{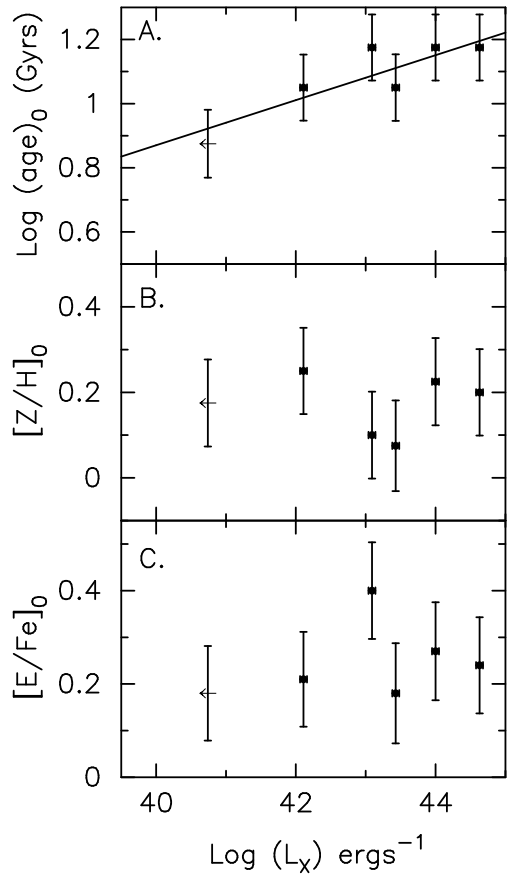}
     } } \end{center}
\caption{The relationship between the X-ray luminosity of the host 
system and central stellar population values.  The error bars indicate
$1\sigma$ errors in X-ray luminosity and the combined random and
systematic errors in the central stellar populations.  Linear fits to
these relationships are given in Table~\ref{cents_fits_table}, and the
significant fit is illustrated.}
\label{env_central}
\end{figure}

In Figure~\ref{env_central} we examine the relationships between the
central stellar populations and their host environment.  Linear
relationships were fitted to the data and are given in the upper
section of Table~\ref{cents_fits_table}. The significant relationship
is illustrated in Figure~\ref{env_central}.  Each panel is discussed
individually below:

{\bf A. Log ($L_X$) -- Log (age)$_0$:} Semi-analytic models of
hierarchical galaxy formation predict galaxies in the field have
younger ages than those in clusters \citep{kauffmann98}: The dense
central regions of the most massive clusters contain the oldest
galaxies at any redshift.

We see that the galaxies in the 5 most X-ray luminous systems have
very similar, old, ages, while the galaxy in the group without
observed extended X-ray emission has the youngest central age.  This
is consistent with the results found for early-type galaxies in
different environments by
\cite{rose94,trager00,kuntschner02,terlevich02,proctor04a,thomas05,bernardi06a,annibali06,smith06}.



{\bf B. Log ($L_X$) -- [Z/H]$_0$:} We observe no relationship between
central metallicity and environment, consistent with observations of
normal early-type galaxies by \cite{bernardi06a} and \cite{smith06},
but in contrast to \cite{rose94,proctor04a,thomas05,annibali06}.

{\bf C. Log ($L_X$) -- [E/Fe]$_0$:} We observe no relationship of
central [E/Fe] with environment, consistent with studies of early-type
galaxies by \cite{thomas05} and \cite{annibali06}.
\cite{smith06} find a significant trend with cluster-centric radius, after removing
the relationship with galaxy mass.  The result found here then
suggests that, at the cluster centre,
the [E/Fe] ratio does not depend on cluster density.

\begin{figure*}	
\begin{center}
     \resizebox{35pc}{!}{
     \rotatebox{-90}{
	\includegraphics{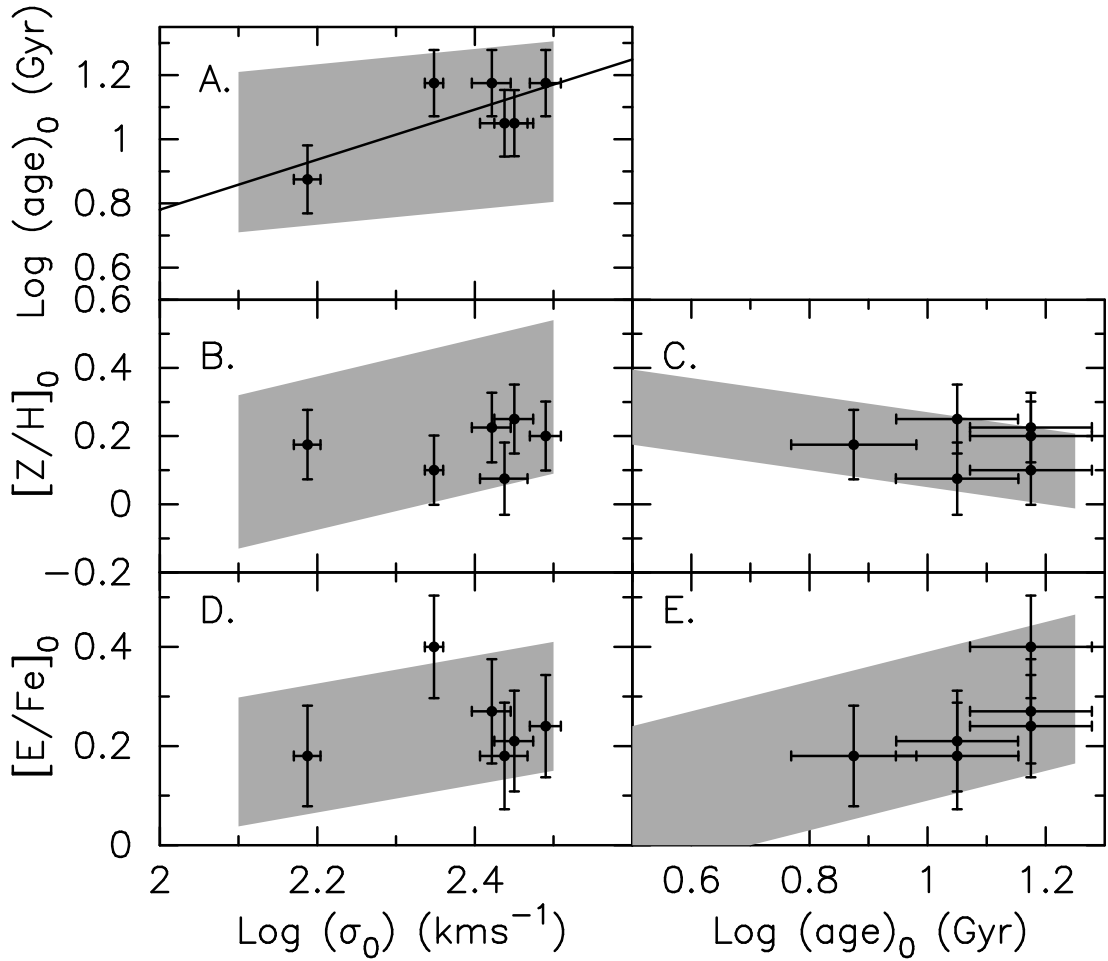}
}  
	}
  \end{center}
\caption{Relationships between central ages, metallicities ([Z/H]), 
$\alpha$-element enhancements ([E/Fe]) and velocity dispersions
($\sigma$) measured in an aperture $a_e/8\times0.5^{\prime\prime}$.
The error bars indicate $1\sigma$ errors in velocity dispersion and
the combined random and systematic errors in the central stellar
populations.  The solid lines indicate straight-line fits to our data
that are statistically significant (all the fits are given in
Table~\ref{cents_fits_table}).  The shaded areas indicate the regions
in which early-type galaxies lie in the relationships from Thomas et
al. (2005; panels A, B, D) and Proctor et al. (2004a; panels C, E).}
\label{central_values}
\end{figure*}

Figure~\ref{central_values} illustrates the relationships between the
central stellar populations themselves.  We also plot the significant
relationships between our parameters (solid lines;
Table~\ref{cents_fits_table}).  For comparison we have also
illustrated the regions in which normal early-type galaxies lie in the
relationships with velocity dispersion in high-density environments
from Thomas et al. (2005; panels A, B, D).  As \cite{thomas05} do not
examine relationships with {\it age}, we also illustrate regions in
which normal early-type galaxies lie in the relationships with age
from Proctor et al. (2004a; panels C, E). \cite{proctor04a} studied
galaxies in the field and loose groups, i.e. low-density environments.
The individual panels are discussed below:

{\bf A. Log $\sigma_0$ -- Log (age)$_0$:} We observe that the 5 most
massive galaxies have very similar masses and central ages, whilst the
least massive galaxy, NGC 3640, has a much lower central age.  As this
is a luminosity-weighted central age, this means that the central
stars of NGC 3640 are either, all 7.5 Gyrs old, or there is a small
fraction of stars younger than 7.5 Gyrs superimposed on an older
stellar population.
Our sample is consistent with the ages measured for early-type
galaxies in high-density environments by \cite{thomas05}.
The trend of older ages in more massive galaxies is consistent with
studies of early-type galaxies by
\cite{trager00,proctor02,proctor04a,thomas05,denicolo05,nelan05,annibali06,mcdermid06}.

{\bf B. Log($\sigma_0$) -- [Z/H]$_0$:} This is the classic
mass-metallicity relationship.  In contrast to the studies of
early-type galaxies by
\cite{mehlert03,proctor04a,denicolo05,nelan05,thomas05,annibali06,mcdermid06}
we observe no relationship of mass with metallicity for the galaxies
in this sample.
However, we do note that our galaxies are consistent with the
relationship followed by normal early-type galaxies in high-density
environments from \cite{thomas05}.  Therefore, the lack of a strong
trend is likely due to the small size of our sample and the small
range in central velocity dispersion probed.

{\bf C. Log (age)$_0$ -- [Z/H]$_0$:} We do not observe a relationship
between metallicity and age.
However, we do note that our galaxies are consistent with the
early-type galaxies studied by \cite{proctor04a}.
\cite{trager00,kuntschner02,mehlert03,denicolo05,thomas05,sanchez06b}
also observe a correlation between these parameters for early-types.
Therefore, the lack of a trend is likely to be a result of the small size
of our sample.



{\bf D. Log ($\sigma_0$) -- [E/Fe]$_0$:} 
We observe no trend of [E/Fe] with mass.  
A positive relationship of [E/Fe] with velocity dispersion is observed
for early-types by
\cite{trager00,terlevich02,proctor02,mehlert03,thomas05,denicolo05,nelan05,annibali06,mcdermid06,sanchez06}.

{\bf E. Log (age)$_0$ -- [E/Fe]$_0$:} We observe no relationship
between age and $\alpha$-element enhancement ratio.  
However, our sample is consistent with the values found for normal
early-type galaxies by \cite{proctor04a}.  
A relationship of age with [E/Fe] ratio is also observed by
\cite{proctor02,mehlert03,proctor04a,denicolo05,thomas05,sanchez06}.
In contrast, \cite{trager00} found [E/Fe] to depend on $\sigma$, not
age for their sample of early-types.




\subsection{Summary}

Our sample of BGGs and BCGs are very uniform in their velocity
dispersions (masses) and stellar populations.
We find one galaxy to be $\leq8$ Gyrs old, while the remaining
galaxies have a small range of central ages of 11-15 Gyrs.

We find that the age of these galaxies correlates with the X-ray
luminosity (i.e. hot gas density) of their host cluster environment.
We also find that the more massive galaxies are older
than the least massive galaxy.


Comparing our results to those obtained for other early-type galaxies,
we find that the stellar populations of these galaxies are consistent
with those of normal early-type galaxies.  This result is
substantiated by \cite{vanderlinden06} who recently studied BCGs in
the Sloan Digital Sky Survey.  They found that, at the same stellar
mass, the stellar populations of non-BCGs and BCGs are very similar,
with the exception of their $\alpha$-element enhancement ratios.  They
found these to be $\sim0.15$ dex higher in BCGs (at our mean velocity
dispersion).  This difference is not visible in our sample, however,
their sample is $\sim100\times$ larger.

These results suggest that BGGs and BCGs have a similar stellar
content and possibly star formation history to other early-type
galaxies.

The lack of a relationship between the central metallicities and
$\alpha$-element enhancement ratios of these galaxies and their host
environment and the uniformity of the metallicities and
$\alpha$-element enhancement ratios is consistent with both a
dissipative collapse and the hierarchical merging predictions of dLB.
The range of ages observed, and their relationship with the host
cluster environment, is only consistent with the predictions of dLB.
It is possible that we can discriminate between these models by
studying the radial gradients in the stellar populations of these
galaxies.







\section{Radial Gradients}
\label{section_grads}

In this section we examine whether the galaxies in our sample show evidence of radial gradients in their stellar populations and velocity dispersion and, if so, in what sense (positive or negative).  
Formation by dissipative collapse predicts strong negative metallicity
gradients (i.e. metallicity decreases with increasing radius) that
correlate with galaxy mass \citep{chiosi02,kawata03}, negative
$\alpha$-element enhancement gradients, and small positive to null age
gradients.  The predictions of dLB lead us to expect negative
metallicity gradients that are shallower in more massive galaxies and
in higher density environments.


The fitted gradients are given in Table~\ref{grads_table} and the
gradients themselves are illustrated in
Figures~\ref{n3557},~\ref{n3640},~\ref{n5044},~\ref{a754},~\ref{a970}
and \ref{a978}.  These are measured within the effective radius at the
observed position angle, i.e. $a_e$, given in Table~\ref{obs_table},
and illustrated as the vertical dotted lines in
Figures~\ref{n3557},~\ref{n3640},~\ref{n5044},~\ref{a754},~\ref{a970}
and \ref{a978}.  We note that if we use the whole galaxy profile
neither our gradients (within the errors) or our results are affected.
Following our method for examining the relationships fitted to this
sample; as a measure of whether or not these gradients are
significantly different from the null hypothesis, we take a gradient
to be significant if it is at least a factor of 3 greater than the
$1\sigma$ error on that gradient (i.e. $\geq3\sigma_{gradient}$).

\subsection{Velocity Dispersion Gradients}

Table~\ref{grads_table} indicates that where a significant velocity
dispersion gradient is observed, it is negative (i.e. velocity
dispersion decreases with increasing radius) but that 5 galaxies in
our sample (the exception being NGC 3557) have velocity dispersion
gradients consistent with zero.
\cite{fisher95a} also found that, with the exception of IC 1101 in 
the Abell 2029 cluster, the velocity dispersion gradients of their
sample of 13 BCGs are negative.  Examining velocity dispersion
gradients to larger radii,
\cite{carter99} found 1 out of their sample of 3 BCGs (NGC 6166 in the
Abell 2199 cluster) to have a positive velocity dispersion gradient.
Rising velocity dispersion profiles have been taken as evidence for
the existence of high mass-to-light ratio components in these galaxies
\citep{dressler79,carter81,carter85}.  These are clearly not present
in this sample over the radii examined here.

\subsection{Age Gradients}

Our sample show small age gradients that are both positive and
negative, but only 2 galaxies have significant age gradients (NGC 3557
and A970\#1).  This is consistent with the results of
\cite{mehlert03} in their sample of early-type galaxies in the Coma
cluster, \cite{sanchez06c} in their high-density environment
early-type galaxy sample and also with the results of \cite{sanchez06}
for early-type galaxies in the field, groups and Virgo cluster
environment.  However, significant, positive, age gradients
(i.e. young central ages and older outer regions) have been observed
by, for example,
\cite{proctor05} in the isolated elliptical galaxy NGC 821, and also
by \cite{sanchez06c} in a larger sample of normal early-type galaxies
in low-density environments.
Positive age gradients suggest recent episodes of secondary star
formation in the centres of galaxies.  
A lack of significant age gradients, together with the, generally, old
central ages of BGGs/BCGs indicates that these galaxies have undergone
few recent episodes of star formation.


\subsection{$\alpha$-element enhancement gradients}

Only 2 galaxies have significant $\alpha$-element enhancement
gradients (NGC 3557 and A978\#1). This is consistent with
\cite{mehlert03} who observe no significant $\alpha$-element
enhancement gradients in their sample of 35 early-type galaxies in
Coma and with the study of the isolated elliptical galaxy NGC 821 by
\cite{proctor05}.  In contrast, \cite{sanchez06} find significant
gradients that are both positive and negative in field, group and
Virgo cluster early-type galaxies.  Current models of the
chemodynamical evolution of galaxies are currently unable to reproduce
galaxies with flat [E/Fe] gradients
(e.g. \citealt{matteucci89,chiappini01,pipino04,pipino06}).

\begin{table}
\begin{center}
\caption{Velocity dispersion and stellar population gradients derived 
for the sample. Errors on the gradients ($1\sigma$) are also given.}
\begin{tabular}{|lcrrc|}
\hline
Galaxy&$\Delta$Log($\sigma$)&$\Delta$Log (age)&$\Delta$[E/Fe]&$\Delta$[Z/H]\\
&/$\Delta$Log$r$&/$\Delta$Log$r$&/$\Delta$Log$r$&/$\Delta$Log$r$\\
&(dex/dex)&(dex/dex)&(dex/dex)&(dex/dex)\\
\hline
NGC 3557&-0.31$\pm$ 0.05& 0.12$\pm$ 0.02& 0.12$\pm$ 0.02&-0.40$\pm$ 0.02\\
NGC 3640&-0.07$\pm$ 0.02& 0.07$\pm$ 0.05& 0.06$\pm$ 0.03&-0.21$\pm$ 0.04\\
NGC 5044&-0.04$\pm$ 0.01& 0.01$\pm$ 0.04& 0.00$\pm$ 0.05&-0.17$\pm$ 0.05\\
A754\#1 &-0.04$\pm$ 0.03&-0.01$\pm$ 0.01& 0.02$\pm$ 0.02&-0.41$\pm$ 0.03\\
A970\#1 &-0.26$\pm$ 0.02&-0.16$\pm$ 0.03& 0.08$\pm$ 0.03&-0.47$\pm$ 0.02\\
A978\#1 &-0.30$\pm$ 0.04& 0.06$\pm$ 0.04& 0.13$\pm$ 0.04&-0.22$\pm$ 0.03\\
\hline
Mean&-0.17$\pm$0.05&0.01$\pm$0.04&0.07$\pm$0.02&-0.31$\pm$0.05\\
\hline
\label{grads_table}
\end{tabular} 
\end{center}
\end{table}

\begin{table}
\begin{center}
\caption{Linear fits to the relationships of the gradients, the significant 
relationships are illustrated in Figures~\ref{zh_grad_mass} and
\ref{zh_grad_rest}.}
\begin{tabular}{|lrll|}
\hline
Relationship (Y - X)&Gradient&Intercept&rms\\
\hline
$\Delta [$Z/H$]/\Delta$Log$r$ - Log($\sigma_{0}$)&-0.72$\pm$ 0.06&  1.37$\pm$ 0.02& 0.12\\
$\Delta [$Z/H$]/\Delta$Log$r$ - $M_K$            &-0.01$\pm$ 0.07& -0.62$\pm$ 0.06& 0.16\\
$\Delta [$Z/H$]/\Delta$Log$r$ - $$[$$Z/H$$]$_{0}$&-1.42$\pm$ 0.04& -0.09$\pm$ 0.02& 0.08\\
$\Delta [$Z/H$]/\Delta$Log$r$ - Log(age)$_{0}$   &-0.66$\pm$ 0.07&  0.35$\pm$ 0.03& 0.15\\
$\Delta [$Z/H$]/\Delta$Log$r$ - Log ($L_X$)      &-0.04$\pm$ 0.01&  1.34$\pm$ 0.01& 0.14\\
\hline
\label{grad_fits_table}
\end{tabular} 
\end{center}
\end{table}

\subsection{Metallicity Gradients}

We find that our sample covers a wide range of metallicity gradients.
The self-consistent numerical GRAPE-SPH simulation of elliptical
galaxy formation by \cite{kobayashi04} found the mean metallicity
gradients for galaxies produced by dissipative collapse (which
includes minor dissipative mergers as the results are similar) and
those produced through major mergers are:
$\Delta[Z/H]/\Delta$log$r\sim-0.30$ (collapse) and $\sim-0.22$ (major
merger).
Individually, the metallicity gradients suggest that three of the
galaxies in our sample have dissipative collapse origins (i.e. NGC
3557, A754\#1, A970\#1 with $\Delta [Z/H]/\Delta$log$r\leq-0.40$) whilst the
other three have a merger origin (i.e. NGC 3640, NGC 5044, A978\#1 with
$\Delta [Z/H]/\Delta$log$r\geq-0.22$).  This range of gradients is
surprising given the small dispersion observed in other BCG
properties.

\begin{figure*}
\begin{center}
     \resizebox{30pc}{!}{
     \rotatebox{-90}{
	\includegraphics{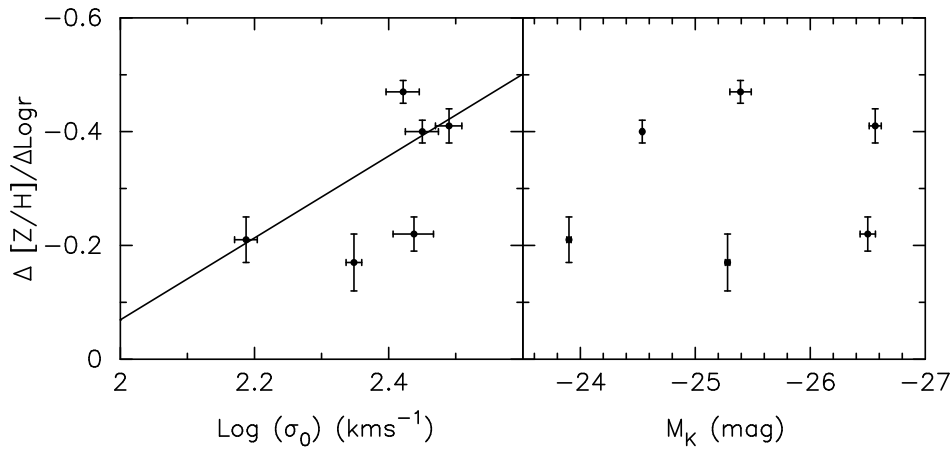}
}  
	}
  \end{center}
\caption{The relationship between metallicity gradient, and mass estimates: 
central velocity dispersion, $\sigma_0$, and $K$-band absolute
magnitude, $M_K$.  The error bars indicate $1\sigma$ errors.  The
solid lines indicate linear fits to our data that are statistically
significant (all fits are given in Table~\ref{grad_fits_table}).}
\label{zh_grad_mass}
\end{figure*}

Any correlation of the gradients with mass is a further means by which
to distinguish between the models of galaxy formation.  In
Figure~\ref{zh_grad_mass} we show metallicity gradients as a function
of proxies for galaxy mass, i.e. velocity dispersion, $\sigma_0$, and
absolute $K$-band magnitude, $M_K$.
We observe a tentative relationship between the metallicity gradient
and the velocity dispersion: more massive galaxies have steeper
gradients.  However, we do not observe a relationship with
$M_K$. \cite{forbes05} found a weak correlation with $\sigma_0$ {\it
and} $M_K$ in the same sense as we observe here for a sample of
early-type galaxies.
\cite{sanchez06} observe a turnaround in this relationship at 
$\sigma_0\sim180$ kms$^{-1}$ (Log($\sigma_0)=2.2$): Normal early-type
galaxies with velocity dispersions less than this have gradients that
steepen with increasing velocity dispersion, as expected from models
of dissipative collapse and galaxies with velocity dispersions more
than this value have shallower gradients, as expected for merger
remnants.  Examining a larger sample of early-type galaxies,
\cite{ogando05} conclude that there is an increase in the scatter of
this relationship above the same velocity dispersion, such that more
massive galaxies show a wider range of metallicity gradients, and
hence a wider range of evolutionary paths, than less massive galaxies.
Our sample lies within the scatter of the data analysed by
\cite{ogando05}, suggesting that the opposite trend we observe is due
to the small size of our sample.



\begin{figure*}
\begin{center}
     \resizebox{40pc}{!}{
     \rotatebox{-90}{
	\includegraphics{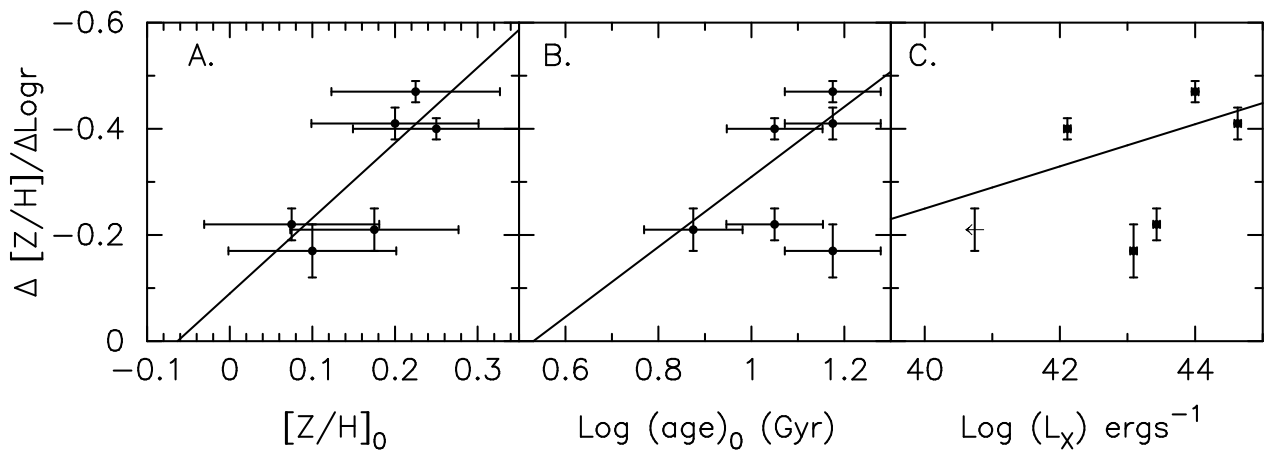}
}  
	}
  \end{center}
\caption{The relationship between metallicity gradient and A: central 
metallicity, B: central age, 
group/cluster X-ray luminosity.  The error bars indicate $1\sigma$
errors.  The solid lines indicate linear fits to our data that are
statistically significant (all the fits are given in
Table~\ref{grad_fits_table}).}
\label{zh_grad_rest}
\end{figure*}

In Figure~\ref{zh_grad_rest} we show the relationships of the
metallicity gradient with other galaxy properties, each panel is
discussed individually below:

{\bf A. [Z/H]$_0$ -- $\Delta$[Z/H]/$\Delta$Log$r$:} We observe a relationship
between the metallicity gradient and the central metallicity such that
galaxies with steeper gradients are more centrally metal-rich.  
\cite{mehlert03} also found a weak hint of a correlation for early-type 
galaxies in the Coma cluster.


{\bf B. Log (age)$_0$ -- $\Delta$[Z/H]/$\Delta$Log$r$:} We find that galaxies with
old central ages have a wide range of metallicity gradients.
Previous research on normal early-type galaxies observed that galaxies
with younger central ages have steeper metallicity gradients
(e.g. \citealt{sanchez06c, sanchez06,kuntschner06}). However, our
galaxies extend to older ages than these samples.
The increased scatter at older ages illustrates the range of
evolutionary histories these massive galaxies must have had.




{\bf C. Log ($L_X$) -- $\Delta$[Z/H]/$\Delta$Log$r$:} We observe a relationship such
that galaxies in more massive clusters have steeper metallicity
gradients, consistent with \cite{sanchez06c}.  



\subsection{Summary}

In contrast to the uniformity of the central stellar populations of
our sample of BGGs and BCGs, we observe them to have a wide range of
metallicity gradients.  We find steeper gradients in galaxies with
higher velocity dispersions, higher central metallicities and in
galaxies with older central ages. The metallicity gradients are also
steeper in galaxies in clusters with high X-ray luminosities.

We find no significant velocity dispersion, age or $\alpha$-element
enhancement ratio gradients in our BGGs or BCGs.

These results are 
consistent with studies of normal early-type galaxies in high-density
environments.  
However, given the homogeneous environment that these galaxies sample,
and the small dispersion of their luminosities, ages and masses, the
range of metallicity gradients we observe is surprising.  Even more
surprising is the fact that the steepest gradients are found in the
most massive galaxies and most X-ray luminous clusters.


We postulate two possible evolutionary scenarios that could produce
the range of gradients we observe and the relationships that they follow:

If simulations of the formation of galaxies and the effects of mergers
on their metallicity gradients are correct
(e.g. \citealt{kobayashi04}), then we expect galaxies that collapsed
at $z>2$ through the assembly of many, large, gas-rich galaxies,
to display steep metallicity gradients, high central metallicities and
old central ages today.  These galaxies would also have shallow age
and $\alpha$-enhancement ratio gradients.  However, if these galaxies
undergo major mergers at $z<2$ then the stellar populations observed
today will depend on whether the merger induced star formation or not:
If the merger did induce star formation then the merger would mix up
the stellar populations, washing out the metallicity gradient, but the
induced star formation would increase the central metallicity again.
In this scenario we would observe shallower metallicity gradients,
average central metallicities and the central ages would depend on
when the last merger occurred.  The galaxies might also have positive
age and $\alpha$-enhancement ratio gradients, depending on when the
collapse occurred.  In contrast, if there was no star formation
induced in the merger, then the merger would wash out the metallicity
gradient without adding a new population of stars.  Therefore, we would
observe shallower metallicity gradients, low central metallicities and
old central ages today, even if the merger was a recent event.  These
galaxies would also show shallow age and $\alpha$-enhancement ratio
gradients.  A combination of these various evolutionary histories are
necessary to explain the relationships we observe with central
metallicity (Figure~\ref{zh_grad_rest} A) and age
(Figure~\ref{zh_grad_rest} B).

However, the models of metallicity gradients have not been thoroughly
tested on observations and it is also possible to explain these
observations through a scenario in which galaxies assemble at high
redshift with different efficiencies: In this scenario all the
galaxies form at high redshift ($>10$ Gyrs ago, $z>2$) and, if the
star formation efficiency remains the same, steep metallicity
gradients are a result of galaxies assembling quickly from many
gas-rich galaxies and passively evolving since, similar to the model
above.  Shallower gradients are then a result of the galaxies
assembling more slowly (at $z\sim2$) from a larger number of less
gas-rich galaxies.  This slower assembly allows the stellar
populations to mix more efficiently, resulting in shallower gradients.
As the assembly takes place relatively quickly at high redshifts this
would also result in small age or $\alpha$-element enhancement ratio
gradients and uniform central stellar populations.  NGC 3640 would
still have had to undergo a relatively recent gaseous merger to
explain its younger central age.  This scenario is similar to the
merger picture outlined above.  The main difference is of timing --
whether the galaxies assembled at $z\sim2$ with little activity since,
or over longer periods of time with either wet or dry major mergers
occurring more recently.

It is also relevant to ask whether it is likely that some of the
mergers occurred without any associated star formation?
Figure 9 of dLB shows that, although the progenitors of BCGs since
$z\sim3$ are gas-poor, some still have non-zero gas fractions,
consistent with observational evidence that most elliptical galaxies
contain some cold gas \citep{morganti06}.  Therefore, it seems
unlikely that even these most massive galaxies would undergo `star
formation-less' mergers, despite their properties (e.g. structure,
rotation; \citealt{fisher95a,brough05,bernardi06b}) being consistent
with purely dissipationless models of galaxy formation
(e.g. \citealt{boylan-kolchin06,cox06,naab06}).  These small gas
fractions are likely to induce small amounts of star formation in the
centres of galaxies that, although not apparent in photometry, would
be visible in their spectra.  We are clearly not observing such star
formation.

The relationship of the metallicity gradients with cluster X-ray
luminosity then suggests that the range of gradients amongst these
galaxies is driven by their host cluster environment.  This is either
due to the lower probability of mergers since $z\sim2$ in the high
velocity dispersion cluster environment, or a higher efficiency of
assembly in the same environments at $z>2$.

It is not possible to determine between these two scenarios with only
6 galaxies.  However, these observations are inconsistent with these
galaxies forming simply through dissipative collapse {\it or} a series
of dissipationless mergers with massive galaxies at recent times, like
the predictions of semi-analytic models of hierarchical galaxy
formation (e.g dLB).  However, they do qualitatively agree with the
results of purely N-body simulations of dark matter evolution in a
Lambda Cold Dark Matter Universe, where the merger rates in massive
cluster haloes are significantly higher than those in group-sized
haloes at redshifts $>2$, but the situation is reversed at redshifts
$<1$ \citep{gottlober01}.



\section{Conclusions}
\label{section_conc}

We have examined the kinematics and stellar populations of a sample of
3 BGGs and 3 BCGs in X-ray groups and clusters.  We have found:


1. The central stellar population of BGGs and BCGs are remarkably
similar to other early-type galaxies in high-density environments,
being indistinguishable in terms of their central metallicities and
$\alpha$-element enhancement ratios.


2. On the other hand we find a wide range of metallicity gradients,
suggesting that these galaxies have had very different assembly
histories.

With a sample of only 6 galaxies, and the scatter observed in these
trends for other early-type galaxies, a larger sample is necessary to
confirm some of the trends we have observed.  However, the similarity
of the stellar populations of BCGs with those of normal early-type
galaxies and the range of metallicity gradients we observe cannot
simply be explained by cosmic variance.  In particular, the range of
metallicity gradients observed is more significant for our small
sample size.

Our observations suggest that BCGs formed at redshifts $z>2$ and that
they must have followed a range of evolutionary histories, dependent
on the density of their host cluster.  Their evolutionary path could
either be a result of the probability of mergers in the cluster
environment since $z\sim2$, or the efficiency of galaxy assembly in
those environments at earlier times ($z>2$).

A lower merger frequency with higher cluster X-ray luminosity is
consistent with the findings from near-infrared photometry of BCGs by
\cite{brough02} and \cite{brough05}, that BCGs in high-mass clusters 
assembled their stellar mass at redshifts $>1$, and have been
passively evolving since, in contrast to BCGs in lower-mass clusters
which appear to still be in the process of merging today.  This is
also consistent with that of N-body simulations \citep{gottlober01}
where more mergers occur for galaxies in clusters than for galaxies in
groups at $z>2$, but that situation reverses at redshifts $<1$.

A higher assembly efficiency is consistent with observations of large
samples of early-type galaxies in various environments by, for
example, \cite{terlevich02}, \cite{thomas05} and \cite{bernardi06a}.
They showed that massive galaxies form earlier and faster than less
massive galaxies.  They suggest that galaxies in low-density
environments form $\sim1-2$ Gyrs after those in high-density
environments, and that these galaxies are more likely to suffer star
formation episodes since $z\sim1$.

BCGs are a special case of the population of Luminous Red Galaxies
(LRGs).  The evolution of this population since redshifts $z\sim1$ has
been studied independently, and the results to date have been
contradictory: \cite{wake06} did not find evidence for evolution
beyond that expected from passive ageing since $z\sim0.6$ in the
luminosity function of their LRGs. They also find their luminosity
functions to be consistent with those of \cite{faber05} at redshifts
$z<0.6$, despite the conclusion of \cite{bell04,faber05} and
\cite{yamada05} that their samples of LRGs have increased in mass by
up to a factor of 4 since $z\sim1$.  \cite{brown06} also found no
evolution since $z\sim1$ for their most massive ($>4L^\star$) LRGs.
Either our merger or assembly alternatives for BCGs are consistent
with these contradictory findings for the evolution of the whole LRG
population.  We also note that the Gemini Deep Deep Survey finds that
the most massive galaxies have already assembled by $z\sim1-2$
\citep{glazebrook04}.

Theoretically, our result that these most massive galaxies cannot all
have formed through simple dissipative collapse, or through a series
of dissipationless mergers with massive galaxies at recent times, is
inconsistent with any single model of galaxy formation.  These results
clearly drive the need for more theoretical understanding of both the
formation and evolution of stellar population gradients and of galaxy
formation itself.

\section*{Acknowledgments}

We would like to thank George Hau and Harald Kuntschner for helpful
comments.  We would also like to thank the anonymous referee for their
very constructive comments. SB, RP, DAF and WJC acknowledge the
funding support of the Australian Research Council.  DJB acknowledges
the support of NASA contract NAS8-39073.  This publication makes use
of data products from the Two Micron All Sky Survey (2MASS) which is a
joint project of the University of Massachusetts and the Infrared
Processing and Analysis Center/California Institute of Technology,
funded by the National Aeronautics and Space Administration and the
National Science Foundation.

\clearpage
\appendix

\section{Central Aperture Indices}
\label{app_centind}
Tables with indices measured in the central $a_e/8$ aperture and their errors per galaxy.

\begin{table*}
\begin{center}
\caption{Value of indices H$\delta_A$ -- Fe4531 and their errors (second line) 
for the central aperture of all galaxies. }
\begin{tabular}{|llllllllllll|}
\hline
Galaxy & H$\delta_A$ & H$\delta_F$
 & CN$_1$ & CN$_2$ & Ca4227 & G4300 & H$\gamma_A$ &
 H$\gamma_F$& Fe4383 & Ca4455 & Fe4531\\
 & \AA\  & \AA\ & mag & mag & \AA\  & \AA\  & \AA\  &
 \AA\  & \AA\  & \AA\  & \AA\ \\
 \hline
NGC 3557&-1.542& 0.088& 0.065& 0.100& 1.210& 4.792&-6.534&-1.847& 5.974& 1.668& 3.260\\& 0.232& 0.166& 0.013& 0.004& 0.064& 0.087& 0.159& 0.069& 0.213& 0.167& 0.104\\
NGC 3640&-0.884& 0.254& 0.026& 0.058& 1.125& 4.837&-6.070&-1.498& 5.425& 1.509& 2.985\\& 0.234& 0.167& 0.013& 0.004& 0.065& 0.090& 0.161& 0.072& 0.202& 0.160& 0.101\\
NGC 5044&-1.889&-0.363& 0.093& 0.126& 1.168& 5.203&-7.860&-3.092& 5.293& 1.395& 2.931\\& 0.236& 0.169& 0.013& 0.004& 0.067& 0.092& 0.163& 0.074& 0.203& 0.160& 0.103\\
A754\#1 &-2.335&-0.070& 0.108& 0.148& 1.233& 5.493&-7.082&-2.224& 5.743& 1.685& 3.429\\& 0.303& 0.210& 0.014& 0.007& 0.108& 0.170& 0.233& 0.129& 0.291& 0.194& 0.187\\
A970\#1 &-2.778&-0.315& 0.124& 0.171& 1.412& 5.684&-7.205&-2.365& 5.595& 1.609& 3.269\\& 0.357& 0.244& 0.014& 0.008& 0.130& 0.207& 0.277& 0.160& 0.334& 0.208& 0.219\\
A978\#1 &-1.479&-0.342& 0.044& 0.086& 1.171& 5.409&-6.330&-1.731& 5.188& 1.950& 3.132\\& 0.315& 0.222& 0.014& 0.007& 0.118& 0.187& 0.253& 0.141& 0.310& 0.200& 0.207\\
\hline
\end{tabular} 
\end{center}
\end{table*}

\begin{table*}
\begin{center}
\caption{Value of indices C4668 -- Fe5782 and their errors (second line) 
for the central aperture of all galaxies. }
\begin{tabular}{|llllllllllll|}
\hline
Galaxy &C4668 & H$\beta$
 &  Fe5015 & Mg$_1$ & Mg$_2$ & Mgb & Fe5270 & Fe5335 &
 Fe5406 & Fe5709 & Fe5782 \\
 & \AA\  & \AA\ 
 & \AA\  & mag & mag & \AA\  & \AA\  & \AA\  &\AA\  & \AA\  & \AA\  \\
 \hline
NGC 3557& 7.332& 1.485& 6.255& 0.134& 0.302& 4.706& 2.857& 2.628& 1.904& 0.774& 1.028\\& 0.126& 0.056& 0.232& 0.004& 0.003& 0.109& 0.111& 0.174& 0.168& 0.054& 0.064\\
NGC 3640& 6.468& 1.899& 5.783& 0.107& 0.260& 4.153& 2.745& 2.420& 1.596& 0.984& 0.870\\& 0.114& 0.054& 0.117& 0.004& 0.003& 0.049& 0.094& 0.083& 0.050& 0.055& 0.048\\
NGC 5044& 6.353&-0.469& 3.346& 0.158& 0.313& 5.595& 2.501& 2.190& 1.557& 0.757& 0.925\\& 0.118& 0.056& 0.165& 0.004& 0.003& 0.054& 0.095& 0.087& 0.056& 0.055& 0.049\\
A754\#1 & 6.722& 1.329& 5.812& 0.226& 0.354& 4.823& 2.598& 2.064& 2.205& 0.805& 0.965\\& 0.250& 0.105& 0.259& 0.004& 0.004& 0.122& 0.141& 0.157& 0.254& 0.085& 0.082\\
A970\#1 & 6.895& 1.386& 5.294& 0.178& 0.307& 5.548& 4.474& 2.499& 1.902& 0.900& 1.105\\& 0.302& 0.123& 0.287& 0.004& 0.004& 0.133& 0.158& 0.177& 0.141& 0.097& 0.093\\
A978\#1 & 5.946& 1.494& 5.698& 0.142& 0.271& 4.552& 3.079& 2.733& 1.810& 0.667& 0.624\\& 0.286& 0.120& 0.303& 0.004& 0.004& 0.137& 0.159& 0.205& 0.171& 0.101& 0.097\\
\hline
\end{tabular} 
\end{center}
\end{table*}

\clearpage

\section{Radial Aperture Indices}
\label{app_apind}
Tables with index values and their errors per aperture per galaxy.

\begin{table*}
\begin{center}
\caption{Value of indices H$\delta_A$ -- Fe4531 and their errors (second line) for apertures from 
$-45^{\prime\prime} - +6^{\prime\prime}$ for NGC 3557. }
\begin{tabular}{|llllllllllll|}
\hline
Radius & H$\delta_A$ & H$\delta_F$
 & CN$_1$ & CN$_2$ & Ca4227 & G4300 & H$\gamma_A$ &
 H$\gamma_F$& Fe4383 & Ca4455 & Fe4531 \\
 ($^{\prime\prime}$) & \AA\  & \AA\  & mag & mag & \AA\  & \AA\  & \AA\  &
 \AA\  & \AA\  & \AA\  & \AA\ \\
 \hline
-45.04&-4.822& 1.469& 0.257& 0.303&-0.003& 2.764&-2.556&-2.109& 0.046& 3.575& 3.389\\& 0.469& 0.275& 0.016& 0.013& 0.211& 0.320& 0.347& 0.225& 0.482& 0.251& 0.312\\
-36.13&-4.314& 0.300& 0.233& 0.267&-0.247& 3.713&-3.662&-1.469& 2.035& 3.014& 2.890\\& 0.493& 0.306& 0.017& 0.014& 0.223& 0.325& 0.369& 0.230& 0.484& 0.259& 0.327\\
-30.88&-3.339& 1.089& 0.176& 0.203&-0.276& 4.521&-5.070&-1.975& 2.817& 2.658& 2.959\\& 0.511& 0.311& 0.017& 0.014& 0.232& 0.342& 0.401& 0.251& 0.509& 0.270& 0.342\\
\end{tabular} 
\end{center}
\end{table*}

\begin{table*}
\begin{center}
\caption{Value of indices C4668 -- Fe5782 and their errors (second line) for apertures from 
$-45^{\prime\prime} - +6^{\prime\prime}$ for NGC 3557. }
\begin{tabular}{|llllllllllll|}
\hline
Radius & C4668 & H$\beta$
 &  Fe5015 & Mg$_1$ & Mg$_2$ & Mgb & Fe5270 & Fe5335 &
 Fe5406 & Fe5709 & Fe5782 \\
 ($^{\prime\prime}$) & \AA\  & \AA\ 
 & \AA\  & mag & mag & \AA\  & \AA\  & \AA\  &\AA\  & \AA\  & \AA\  \\
 \hline
-45.04& 5.419& 5.804& 3.866& 0.071& 0.200& 2.932& 2.351& 1.727&-0.677&-0.388& 0.217\\& 0.462& 0.154& 0.364& 0.006& 0.005& 0.172& 0.199& 0.220& 0.171& 0.148& 0.124\\
-36.13& 6.488& 4.734& 4.439& 0.095& 0.202& 3.054& 2.019& 2.098& 0.066&-0.314& 0.217\\& 0.466& 0.164& 0.380& 0.006& 0.005& 0.177& 0.208& 0.226& 0.171& 0.151& 0.126\\
-30.88& 6.512& 3.997& 4.917& 0.091& 0.227& 3.359& 2.091& 2.028& 0.361& 0.209& 0.345\\& 0.491& 0.177& 0.401& 0.006& 0.006& 0.188& 0.218& 0.240& 0.180& 0.155& 0.132\\
\end{tabular} 
\end{center}
\end{table*}

\begin{table*}
\begin{center}
\caption{Value of indices H$\delta_A$ -- Fe4531 and their errors (second line) for apertures from 
$+6^{\prime\prime} - +45^{\prime\prime}$. The apertures from +8 -- +45$^{\prime\prime}$ 
are contaminated by another galaxy in the slit.}
\begin{tabular}{|llllllllllll|}
\hline
Radius & H$\delta_A$ & H$\delta_F$
 & CN$_1$ & CN$_2$ & Ca4227 & G4300 & H$\gamma_A$ &
 H$\gamma_F$& Fe4383 & Ca4455 & Fe4531 \\
 ($^{\prime\prime}$) & \AA\  & \AA\ & mag & mag & \AA\  & \AA\  & \AA\  &
 \AA\  & \AA\  & \AA\ & \AA\ \\
 \hline
  6.57&-1.439& 0.030& 0.042& 0.072& 1.108& 4.786&-6.180&-1.855& 5.321& 1.411& 2.868\\& 0.354& 0.244& 0.015& 0.009& 0.133& 0.219& 0.280& 0.161& 0.337& 0.212& 0.223\\
  7.59&-1.604& 0.045& 0.047& 0.068& 1.114& 4.300&-5.690&-1.500& 5.042& 1.453& 2.970\\& 0.378& 0.259& 0.015& 0.010& 0.144& 0.242& 0.300& 0.173& 0.362& 0.222& 0.242\\
  8.61&-1.792& 0.013& 0.052& 0.079& 1.154& 4.754&-5.950&-1.654& 5.181& 1.568& 2.901\\& 0.403& 0.275& 0.015& 0.011& 0.155& 0.261& 0.325& 0.190& 0.389& 0.233& 0.264\\
\end{tabular} 
\end{center}
\end{table*}

\begin{table*}
\begin{center}
\caption{Value of indices C4668 -- Fe5782 and their errors (second line) for apertures from 
$+6^{\prime\prime} - +45^{\prime\prime}$.  The apertures from +8 -- +45$^{\prime\prime}$ 
are contaminated by another galaxy in the slit.}
\begin{tabular}{|llllllllllll|}
\hline
Radius & C4668 & H$\beta$
 &  Fe5015 & Mg$_1$ & Mg$_2$ & Mgb & Fe5270 & Fe5335 &
 Fe5406 & Fe5709 & Fe5782 \\
 ($^{\prime\prime}$) & \AA\  & \AA\ 
 & \AA\  & mag & mag & \AA\  & \AA\  & \AA\  &\AA\  & \AA\  & \AA\  \\
 \hline
  6.57& 6.631& 1.724& 5.574& 0.124& 0.278& 4.217& 2.470& 2.071& 1.614& 0.684& 0.709\\& 0.303& 0.121& 0.278& 0.004& 0.004& 0.124& 0.155& 0.174& 0.135& 0.111& 0.088\\
  7.59& 6.003& 1.550& 4.959& 0.119& 0.276& 4.287& 2.491& 2.012& 1.427& 0.771& 0.654\\& 0.335& 0.133& 0.290& 0.005& 0.004& 0.130& 0.163& 0.175& 0.146& 0.117& 0.095\\
  8.61& 6.563& 1.793& 5.784& 0.120& 0.279& 4.385& 2.718& 2.172& 1.544& 0.540& 0.699\\& 0.364& 0.145& 0.338& 0.005& 0.004& 0.147& 0.176& 0.196& 0.143& 0.126& 0.103\\
\end{tabular} 
\end{center}
\end{table*}

\begin{table*}
\begin{center}
\caption{Value of indices H$\delta_A$ -- Fe4531 and their errors (second line) for apertures from 
$-45^{\prime\prime} - +10^{\prime\prime}$ for NGC 3640. }
\begin{tabular}{|llllllllllll|}
\hline
Radius & H$\delta_A$ & H$\delta_F$
 & CN$_1$ & CN$_2$ & Ca4227 & G4300 & H$\gamma_A$ &
 H$\gamma_F$& Fe4383 & Ca4455 & Fe4531\\
  ($^{\prime\prime}$) & \AA\  & \AA\ & mag & mag & \AA\  & \AA\  & \AA\  &
 \AA\  & \AA\  & \AA\  & \AA\  \\
 \hline
-44.9& 2.659& 0.573&-0.087&-0.063& 1.331& 3.467&-3.864&-0.993& 2.400& 1.799& 2.977\\& 0.384& 0.281& 0.016& 0.011& 0.168& 0.291& 0.340& 0.206& 0.451& 0.251& 0.300\\
-31.5& 2.830& 0.673&-0.099&-0.088& 0.965& 4.190&-4.648&-0.601& 3.682& 1.275& 2.946\\& 0.436& 0.322& 0.017& 0.014& 0.200& 0.331& 0.388& 0.230& 0.487& 0.276& 0.334\\
-25.5& 1.495& 0.785&-0.079&-0.071& 1.122& 3.836&-4.754&-0.798& 3.705& 1.177& 3.328\\& 0.457& 0.324& 0.017& 0.014& 0.194& 0.334& 0.392& 0.234& 0.490& 0.276& 0.331\\
\end{tabular} 
\end{center}
\end{table*}

\begin{table*}
\begin{center}
\caption{Value of indices C4668 -- Fe5782 and their errors (second line) for apertures from 
$-45^{\prime\prime} - +10^{\prime\prime}$ for NGC 3640.}
\begin{tabular}{|llllllllllll|}
\hline
Radius & C4668 & H$\beta$
 &  Fe5015 & Mg$_1$ & Mg$_2$ & Mgb & Fe5270 & Fe5335 &
 Fe5406 & Fe5709 & Fe5782 \\
  ($^{\prime\prime}$) & \AA\  & \AA\ 
 & \AA\  & mag & mag & \AA\  & \AA\  & \AA\  &\AA\  & \AA\  & \AA\  \\
 \hline
-44.9& 4.612& 4.792& 2.466& 0.083& 0.222& 4.057& 2.484& 1.566& 1.735& 1.031& 1.084\\& 0.432& 0.150& 0.358& 0.005& 0.005& 0.157& 0.192& 0.213& 0.150& 0.126& 0.115\\
-31.5& 5.037& 3.135& 3.909& 0.067& 0.212& 3.368& 2.122& 1.506& 1.442& 0.721& 0.565\\& 0.478& 0.176& 0.393& 0.006& 0.006& 0.181& 0.215& 0.236& 0.167& 0.140& 0.130\\
-25.5& 4.717& 2.707& 3.760& 0.079& 0.216& 3.757& 2.796& 1.967& 1.519& 0.723& 0.646\\& 0.480& 0.179& 0.395& 0.006& 0.006& 0.177& 0.210& 0.231& 0.167& 0.140& 0.129\\
\end{tabular} 
\end{center}
\end{table*}

\begin{table*}
\begin{center}
\caption{Value of indices H$\delta_A$ -- Fe4531 and their errors (second line) for apertures from 
$+10^{\prime\prime} - +43^{\prime\prime}$ for NGC 3640. }
\begin{tabular}{|llllllllllll|}
\hline
Radius & H$\delta_A$ & H$\delta_F$
 & CN$_1$ & CN$_2$ & Ca4227 & G4300 & H$\gamma_A$ &
 H$\gamma_F$& Fe4383 & Ca4455 & Fe4531\\
  ($^{\prime\prime}$) & \AA\  & \AA\ & mag & mag & \AA\  & \AA\  & \AA\  &
 \AA\  & \AA\  & \AA\  & \AA\  \\
 \hline
 10.7&-0.384& 0.547&-0.019& 0.008& 1.115& 4.938&-5.001&-1.099& 4.377& 1.333& 2.678\\& 0.455& 0.311& 0.016& 0.013& 0.185& 0.306& 0.377& 0.227& 0.457& 0.264& 0.322\\
 11.8&-0.208& 0.554&-0.038&-0.013& 0.992& 4.747&-5.169&-1.267& 4.236& 1.237& 2.701\\& 0.451& 0.310& 0.016& 0.013& 0.188& 0.307& 0.377& 0.229& 0.462& 0.266& 0.321\\
 13.0& 0.866& 0.643&-0.056&-0.033& 1.056& 4.215&-4.905&-0.933& 4.813& 0.950& 2.683\\& 0.449& 0.314& 0.016& 0.013& 0.189& 0.317& 0.381& 0.227& 0.462& 0.272& 0.328\\
 14.5& 0.133& 0.056&-0.041&-0.012& 1.269& 4.832&-5.397&-1.194& 4.403& 1.048& 2.191\\& 0.450& 0.317& 0.016& 0.013& 0.184& 0.306& 0.379& 0.228& 0.460& 0.268& 0.327\\
 15.5& 2.021& 1.273&-0.085&-0.054& 1.487& 4.900&-5.573&-0.941& 4.822& 1.536& 2.297\\& 0.673& 0.469& 0.022& 0.020& 0.297& 0.519& 0.629& 0.380& 0.755& 0.398& 0.556\\
 17.1& 0.752& 0.555&-0.037&-0.011& 1.130& 4.732&-5.543&-1.423& 4.416& 1.165& 2.403\\& 0.419& 0.295& 0.016& 0.012& 0.172& 0.287& 0.358& 0.215& 0.434& 0.254& 0.304\\
 19.9& 0.211& 0.648&-0.032&-0.010& 1.212& 4.452&-5.225&-0.972& 4.759& 1.439& 2.954\\& 0.429& 0.295& 0.016& 0.012& 0.175& 0.293& 0.360& 0.213& 0.437& 0.254& 0.303\\
 23.9& 0.542& 0.539&-0.035&-0.008& 1.374& 4.266&-4.803&-0.803& 4.295& 1.177& 2.388\\& 0.428& 0.300& 0.016& 0.012& 0.171& 0.298& 0.358& 0.212& 0.439& 0.259& 0.309\\
 29.9& 1.374& 1.144&-0.056&-0.044& 0.565& 4.018&-4.812&-0.773& 3.787& 0.840& 2.558\\& 0.425& 0.295& 0.016& 0.012& 0.186& 0.300& 0.359& 0.213& 0.446& 0.263& 0.315\\
 43.3& 2.885&-0.008&-0.105&-0.111& 0.992& 4.184&-5.539&-1.075& 4.341& 0.832& 2.210\\& 0.392& 0.299& 0.016& 0.012& 0.172& 0.289& 0.354& 0.208& 0.434& 0.259& 0.309\\
\hline
\label{appendix_indices_n3640_3}
\end{tabular} 
\end{center}
\end{table*}

\begin{table*}
\begin{center}
\caption{Value of indices C4668 -- Fe5782 and their errors (second line) for apertures from 
$+10^{\prime\prime} - +43^{\prime\prime}$ for NGC 3640.}
\begin{tabular}{|llllllllllll|}
\hline
Radius & C4668 & H$\beta$
 &  Fe5015 & Mg$_1$ & Mg$_2$ & Mgb & Fe5270 & Fe5335 &
 Fe5406 & Fe5709 & Fe5782 \\
  ($^{\prime\prime}$) & \AA\  & \AA\ 
 & \AA\  & mag & mag & \AA\  & \AA\  & \AA\  &\AA\  & \AA\  & \AA\  \\
 \hline
 10.7& 4.665& 2.031& 4.897& 0.085& 0.222& 3.616& 2.307& 2.091& 1.396& 0.914& 0.643\\& 0.458& 0.178& 0.383& 0.006& 0.005& 0.177& 0.211& 0.230& 0.167& 0.139& 0.130\\
 11.8& 4.708& 1.709& 4.791& 0.086& 0.231& 3.911& 2.487& 2.149& 1.341& 0.898& 0.485\\& 0.461& 0.181& 0.384& 0.006& 0.005& 0.177& 0.210& 0.229& 0.167& 0.139& 0.130\\
 13.0& 4.724& 2.046& 4.588& 0.081& 0.228& 3.890& 2.519& 2.109& 1.767& 1.160& 0.752\\& 0.469& 0.181& 0.391& 0.006& 0.006& 0.179& 0.213& 0.233& 0.167& 0.140& 0.130\\
\end{tabular} 
\end{center}
\end{table*}

\begin{table*}
\begin{center}
\caption{Value of indices H$\delta_A$ -- Fe4531 and their errors (second line) for apertures from 
$-50^{\prime\prime} - 0^{\prime\prime}$ for NGC 5044.}
\begin{tabular}{|llllllllllll|}
\hline
Radius & H$\delta_A$ & H$\delta_F$
 & CN$_1$ & CN$_2$ & Ca4227 & G4300 & H$\gamma_A$ &
 H$\gamma_F$& Fe4383 & Ca4455 & Fe4531\\
 ($^{\prime\prime}$) & \AA\  & \AA\ & mag & mag & \AA\  & \AA\  & \AA\  &
 \AA\  & \AA\  & \AA\  & \AA\  \\
 \hline
-49.6&-1.714& 1.579& 0.049& 0.048& 1.181& 3.112&-3.993&-0.035& 6.195& 2.221& 2.366 \\& 0.447& 0.283& 0.016& 0.013& 0.182& 0.323& 0.368& 0.213& 0.446& 0.255& 0.329 \\
-42.8& 0.079&-0.347& 0.042& 0.064& 0.998& 4.447&-4.980&-1.295& 4.661& 0.960& 1.950 \\& 0.465& 0.327& 0.017& 0.014& 0.198& 0.332& 0.396& 0.241& 0.488& 0.281& 0.351 \\
-38.2& 1.824& 1.570& 0.005& 0.037& 1.663& 3.782&-4.600&-0.882& 5.405& 0.723& 3.135 \\& 0.459& 0.307& 0.017& 0.014& 0.193& 0.338& 0.392& 0.234& 0.477& 0.282& 0.341 \\
\end{tabular} 
\end{center}
\end{table*}

\begin{table*}
\begin{center}
\caption{Value of indices C4668 -- Fe5782 and their errors (second line) for apertures from 
$-50^{\prime\prime} - 0^{\prime\prime}$ for NGC 5044.}
\begin{tabular}{|llllllllllll|}
\hline
Radius & C4668 & H$\beta$
 &  Fe5015 & Mg$_1$ & Mg$_2$ & Mgb & Fe5270 & Fe5335 &
 Fe5406 & Fe5709 & Fe5782 \\
 ($^{\prime\prime}$) & \AA\  & \AA\ 
 & \AA\  & mag & mag & \AA\  & \AA\  & \AA\  &\AA\  & \AA\  & \AA\  \\
 \hline
-49.6&4.856& 0.656& 2.929& 0.126& 0.256& 4.089& 2.986& 0.720& 2.039& 1.193& 1.567 \\&0.465& 0.193& 0.393& 0.006& 0.005& 0.178& 0.208& 0.234& 0.164& 0.157& 0.122\\
-42.8&3.715& 0.565& 4.931& 0.135& 0.287& 4.911& 2.851& 1.184& 1.464& 1.218& 1.427 \\&0.495& 0.203& 0.407& 0.006& 0.006& 0.187& 0.219& 0.246& 0.184& 0.162& 0.130\\
-38.2&2.629& 0.825& 5.309& 0.151& 0.304& 4.422& 2.524& 1.199& 1.460& 0.745& 1.478 \\&0.497& 0.200& 0.400& 0.006& 0.006& 0.186& 0.218& 0.241& 0.175& 0.164& 0.127\\
\end{tabular} 
\end{center}
\end{table*}

\begin{table*}
\begin{center}
\caption{Value of indices H$\delta_A$ -- Fe4531 and their errors (second line) for apertures from 
$0^{\prime\prime} - +50^{\prime\prime}$ for NGC 5044.}
\begin{tabular}{|llllllllllll|}
\hline
Radius & H$\delta_A$ & H$\delta_F$
 & CN$_1$ & CN$_2$ & Ca4227 & G4300 & H$\gamma_A$ &
 H$\gamma_F$& Fe4383 & Ca4455 & Fe4531\\
 ($^{\prime\prime}$) & \AA\  & \AA\ & mag & mag & \AA\  & \AA\  & \AA\  &
 \AA\  & \AA\  & \AA\  & \AA\  \\
 \hline
  0.4&-2.142&-0.418& 0.110& 0.146& 1.198& 5.350&-8.450&-3.455& 5.289& 1.338& 2.879 \\& 0.284& 0.192& 0.014& 0.006& 0.091& 0.133& 0.201& 0.106& 0.241& 0.173& 0.146 \\
  1.5&-1.832&-0.361& 0.089& 0.121& 1.191& 5.132&-7.558&-2.945& 5.131& 1.504& 2.900 \\& 0.292& 0.198& 0.014& 0.006& 0.095& 0.141& 0.207& 0.111& 0.250& 0.176& 0.153 \\
  2.5&-1.751&-0.202& 0.088& 0.120& 1.163& 5.262&-7.747&-2.978& 5.539& 1.434& 3.009 \\& 0.306& 0.206& 0.014& 0.006& 0.103& 0.154& 0.222& 0.122& 0.263& 0.181& 0.166 \\
\end{tabular} 
\end{center}
\end{table*}

\begin{table*}
\begin{center}
\caption{Value of indices C4668 -- Fe5782 and their errors (second line) for apertures from 
$0^{\prime\prime} - +50^{\prime\prime}$ for NGC 5044.}
\begin{tabular}{|lllllllllllllllllllllll|}
\hline
Radius & C4668 & H$\beta$
 &  Fe5015 & Mg$_1$ & Mg$_2$ & Mgb & Fe5270 & Fe5335 &
 Fe5406 & Fe5709 & Fe5782 \\
 ($^{\prime\prime}$) & \AA\  & \AA\ 
 & \AA\  & mag & mag & \AA\  & \AA\  & \AA\  &\AA\  & \AA\  & \AA\  \\
 \hline
  0.4&6.188&-1.171& 1.744& 0.159& 0.304& 5.798& 2.488& 2.172& 1.446& 0.783& 0.922 \\&0.181& 0.100& 0.173& 0.004& 0.003& 0.080& 0.114& 0.111& 0.087& 0.109& 0.068\\
  1.5&6.360&-0.173& 3.892& 0.158& 0.316& 5.527& 2.486& 2.255& 1.657& 0.725& 0.886 \\&0.191& 0.102& 0.218& 0.004& 0.004& 0.080& 0.116& 0.112& 0.087& 0.110& 0.069\\
  2.5&6.581&-0.404& 4.442& 0.156& 0.313& 5.651& 2.549& 2.282& 1.771& 0.726& 0.988 \\&0.211& 0.111& 0.193& 0.004& 0.004& 0.089& 0.122& 0.122& 0.096& 0.113& 0.073\\
\end{tabular} 
\end{center}
\end{table*}

\begin{table*}
\begin{center}
\caption{Value of indices H$\delta_A$ -- Fe4531 and their errors (second line) for apertures from 
$-21^{\prime\prime} - +21^{\prime\prime}$ for A754\#1.}
\begin{tabular}{|llllllllllll|}
\hline
Radius & H$\delta_A$ & H$\delta_F$ & CN$_1$ & CN$_2$ & Ca4227 & G4300 & H$\gamma_A$ & H$\gamma_F$& Fe4383 & Ca4455 & Fe4531 \\
 ($^{\prime\prime}$) & \AA\  & \AA\ & mag & mag & \AA\  & \AA\  & \AA\  & \AA\  & \AA\  & \AA\  & \AA\  \\
 \hline
-20.8& 1.520& 2.012& 0.001& 0.015& 0.983& 5.376&-3.985&-1.589& 6.312& 1.633& 4.186\\& 0.351& 0.240& 0.010& 0.013& 0.198& 0.355& 0.391& 0.248& 0.614& 0.372& 0.319\\
-10.7& 0.607& 0.755& 0.029& 0.044& 0.361& 5.474&-4.681&-0.817& 5.469& 0.829& 2.765\\& 0.382& 0.269& 0.010& 0.013& 0.202& 0.356& 0.393& 0.242& 0.606& 0.333& 0.320\\
 -6.5&-1.870& 0.462& 0.042& 0.064& 1.215& 5.159&-5.352&-1.605& 5.346& 0.811& 3.085\\& 0.399& 0.264& 0.010& 0.013& 0.190& 0.350& 0.390& 0.244& 0.604& 0.332& 0.315\\
\end{tabular} 
\end{center}
\end{table*}

\begin{table*}
\begin{center}
\caption{Value of indices C4668 -- Fe5782 and their errors (second line) for apertures from 
$-21^{\prime\prime} - +21^{\prime\prime}$ for A754\#1.}
\begin{tabular}{|llllllllllll|}
\hline
Radius & C4668 & H$\beta$ &  Fe5015 & Mg$_1$ & Mg$_2$ & Mgb & Fe5270 & Fe5335 & Fe5406 & Fe5709 & Fe5782 \\
 ($^{\prime\prime}$) & \AA\  & \AA\  & \AA\  & mag & mag & \AA\  & \AA\  & \AA\  &\AA\  & \AA\  & \AA\  \\
 \hline
-20.8& 7.707& 0.234& 5.863& 0.173& 0.179& 2.582&-0.442&-0.090& 2.215& 0.205& 1.076\\& 0.484& 0.200& 0.502& 0.004& 0.008& 0.182& 0.591& 0.377& 0.273& 0.148& 0.132\\
-10.7& 8.033& 1.124& 4.149& 0.129& 0.181& 2.476& 0.320& 0.897& 1.098& 0.783& 0.565\\& 0.483& 0.193& 0.503& 0.004& 0.008& 0.186& 0.595& 0.381& 0.255& 0.146& 0.133\\
 -6.5& 7.737& 1.213& 4.187& 0.135& 0.215& 3.491& 1.147& 0.746& 2.119& 0.590& 0.777\\& 0.479& 0.191& 0.507& 0.004& 0.008& 0.187& 0.598& 0.386& 0.253& 0.145& 0.129\\
\end{tabular} 
\end{center}
\end{table*}

\begin{table*}
\begin{center}
\caption{Value of indices H$\delta_A$ -- Fe4531 and their errors (second line) for apertures from 
$-7^{\prime\prime} - +7^{\prime\prime}$ for A970\#1.}
\begin{tabular}{|llllllllllll|}
\hline
Radius & H$\delta_A$ & H$\delta_F$ & CN$_1$ & CN$_2$ & Ca4227 & G4300 & H$\gamma_A$ & H$\gamma_F$& Fe4383 & Ca4455 & Fe4531 \\
 ($^{\prime\prime}$) & \AA\  & \AA\ & mag & mag & \AA\  & \AA\  & \AA\  & \AA\  & \AA\  & \AA\  & \AA\  \\
 \hline
-7.2&-0.106& 0.901&-0.064&-0.036& 0.493& 5.048&-6.547&-1.822& 3.755& 1.451& 1.866\\& 0.688& 0.459& 0.021& 0.020& 0.330& 0.538& 0.642& 0.397& 0.832& 0.436& 0.608\\
-4.2&-0.250&-0.067& 0.052& 0.061& 1.347& 4.792&-6.452&-2.205& 3.728& 1.362& 2.046\\& 0.680& 0.479& 0.021& 0.020& 0.283& 0.515& 0.604& 0.380& 0.786& 0.409& 0.579\\
-2.6&-0.750& 0.658& 0.025& 0.061& 1.100& 6.406&-6.535&-2.182& 5.179& 1.767& 2.443\\& 0.687& 0.460& 0.021& 0.019& 0.296& 0.493& 0.611& 0.383& 0.763& 0.405& 0.570\\
-1.6&-2.313& 0.343& 0.079& 0.120& 1.416& 5.671&-7.326&-2.582& 5.015& 1.361& 2.436\\& 0.555& 0.361& 0.018& 0.015& 0.218& 0.371& 0.458& 0.283& 0.601& 0.334& 0.446\\
-0.6&-2.499&-0.105& 0.117& 0.163& 1.326& 5.719&-6.902&-2.316& 5.394& 1.707& 3.416\\& 0.426& 0.280& 0.015& 0.010& 0.154& 0.266& 0.330& 0.194& 0.472& 0.277& 0.347\\
 0.4&-3.047&-0.125& 0.125& 0.169& 1.449& 5.658&-7.018&-2.352& 5.601& 1.590& 3.201\\& 0.378& 0.248& 0.014& 0.009& 0.132& 0.231& 0.289& 0.166& 0.437& 0.265& 0.325\\
 1.5&-2.202& 0.354& 0.097& 0.143& 1.259& 5.298&-6.452&-2.282& 5.530& 1.697& 3.632\\& 0.426& 0.278& 0.015& 0.011& 0.162& 0.281& 0.347& 0.207& 0.493& 0.288& 0.369\\
 2.5&-1.687& 0.291& 0.064& 0.101& 1.108& 4.964&-5.428&-1.684& 5.070& 1.382& 2.144\\& 0.549& 0.364& 0.018& 0.015& 0.228& 0.397& 0.467& 0.285& 0.629& 0.350& 0.479\\
 4.0&-0.789& 0.720& 0.035& 0.077& 1.038& 5.252&-6.343&-1.604& 5.197& 1.428& 3.036\\& 0.566& 0.372& 0.018& 0.016& 0.243& 0.414& 0.500& 0.302& 0.650& 0.358& 0.486\\
 7.1&-1.090& 0.585&-0.037&-0.003& 0.475& 4.792&-5.501&-1.742& 2.989& 2.309&-0.552\\& 0.632& 0.419& 0.019& 0.018& 0.284& 0.482& 0.563& 0.353& 0.760& 0.387& 0.571\\
\hline
\label{appendix_indices_a970}
\end{tabular} 
\end{center}
\end{table*}

\begin{table*}
\begin{center}
\caption{Value of indices C4668 -- Fe5782 and their errors (second line) for apertures from 
$-7^{\prime\prime} - +7^{\prime\prime}$ for A970\#1.}
\begin{tabular}{|llllllllllll|}
\hline
Radius & C4668 & H$\beta$ &  Fe5015 & Mg$_1$ & Mg$_2$ & Mgb & Fe5270 & Fe5335 & Fe5406 & Fe5709 & Fe5782 \\
 ($^{\prime\prime}$) & \AA\  & \AA\  & \AA\  & mag & mag & \AA\  & \AA\  & \AA\  &\AA\  & \AA\  & \AA\  \\
 \hline
-7.2& 2.063& 2.048& 2.444& 0.117& 0.197& 3.157& 0.294& 1.119& 0.663& 1.073& 1.127\\& 0.878& 0.315& 0.707& 0.007& 0.009& 0.317& 0.372& 0.418& 0.358& 0.239& 0.245\\
-4.2& 5.382& 2.525& 4.847& 0.124& 0.236& 3.904& 0.328& 1.450& 1.221& 1.078& 1.253\\& 0.811& 0.288& 0.632& 0.007& 0.008& 0.289& 0.345& 0.383& 0.334& 0.220& 0.225\\
-2.6& 4.970& 1.809& 4.772& 0.145& 0.267& 4.653& 1.135& 1.369& 1.644& 1.197& 0.379\\& 0.815& 0.293& 0.629& 0.007& 0.008& 0.284& 0.339& 0.381& 0.333& 0.214& 0.227\\
-1.6& 5.635& 1.302& 5.672& 0.167& 0.288& 5.101& 1.303& 2.284& 1.530& 0.949& 0.693\\& 0.625& 0.217& 0.466& 0.006& 0.006& 0.204& 0.247& 0.278& 0.273& 0.154& 0.173\\
-0.6& 6.604& 1.576& 5.293& 0.179& 0.309& 5.267& 3.432& 2.272& 1.802& 0.839& 1.062\\& 0.490& 0.152& 0.311& 0.005& 0.005& 0.182& 0.179& 0.203& 0.232& 0.110& 0.147\\
 0.4& 7.216& 1.268& 5.576& 0.175& 0.301& 5.620& 4.549& 2.486& 1.993& 0.844& 1.017\\& 0.455& 0.140& 0.300& 0.005& 0.005& 0.140& 0.193& 0.244& 0.286& 0.106& 0.138\\
 1.5& 6.425& 1.349& 5.517& 0.166& 0.290& 5.344& 2.392& 2.351& 2.108& 1.002& 0.990\\& 0.530& 0.180& 0.410& 0.006& 0.005& 0.179& 0.218& 0.254& 0.267& 0.143& 0.165\\
 2.5& 5.839& 1.549& 4.679& 0.161& 0.275& 4.947& 1.804& 1.893& 2.027& 0.515& 1.064\\& 0.683& 0.248& 0.547& 0.006& 0.007& 0.256& 0.296& 0.344& 0.322& 0.204& 0.212\\
 4.0& 5.041& 2.040& 4.176& 0.142& 0.236& 4.368& 1.747& 1.392& 1.198& 1.437& 1.050\\& 0.707& 0.255& 0.555& 0.006& 0.007& 0.253& 0.300& 0.348& 0.313& 0.201& 0.214\\
 7.1& 1.315& 0.993& 4.574& 0.119& 0.228& 3.741& 2.687& 1.448& 0.918& 0.351& 0.927\\& 0.802& 0.290& 0.621& 0.007& 0.008& 0.289& 0.321& 0.387& 0.339& 0.230& 0.233\\
\hline
\label{appendix_indices_a970_2}
\end{tabular} 
\end{center}
\end{table*}

\begin{table*}
\begin{center}
\caption{Value of indices H$\delta_A$ -- Fe4531 and their errors (second line) for apertures from 
$-14^{\prime\prime} - +13^{\prime\prime}$ for A978\#1.}
\begin{tabular}{|llllllllllll|}
\hline
Radius & H$\delta_A$ & H$\delta_F$ & CN$_1$ & CN$_2$ & Ca4227 & G4300 & H$\gamma_A$ & H$\gamma_F$& Fe4383 & Ca4455 & Fe4531 \\
 ($^{\prime\prime}$) & \AA\  & \AA\ & mag & mag & \AA\  & \AA\  & \AA\  & \AA\  & \AA\  & \AA\  & \AA\  \\
 \hline
-13.6&-0.189&-0.077&-0.036& 0.002& 1.120& 4.049&-5.546&-2.409& 3.442& 1.059& 2.970\\& 0.430& 0.311& 0.016& 0.012& 0.186& 0.334& 0.380& 0.245& 0.480& 0.265& 0.334\\
 -7.5&-1.308& 0.321& 0.011& 0.047& 0.866& 4.802&-6.797&-1.890& 4.561& 1.395& 3.379\\& 0.427& 0.296& 0.016& 0.011& 0.179& 0.313& 0.369& 0.227& 0.449& 0.260& 0.320\\
 -4.4&-1.148&-0.264& 0.023& 0.049& 1.032& 4.646&-6.089&-1.882& 4.448& 1.411& 3.134\\& 0.358& 0.259& 0.015& 0.009& 0.142& 0.256& 0.299& 0.185& 0.369& 0.220& 0.263\\
\end{tabular} 
\end{center}
\end{table*}

\begin{table*}
\begin{center}
\caption{Value of indices C4668 -- Fe5782 and their errors (second line) for apertures from 
$-14^{\prime\prime} - +13^{\prime\prime}$ for A978\#1.}
\begin{tabular}{|llllllllllll|}
\hline
Radius & C4668 & H$\beta$ &  Fe5015 & Mg$_1$ & Mg$_2$ & Mgb & Fe5270 & Fe5335 & Fe5406 & Fe5709 & Fe5782 \\
 ($^{\prime\prime}$) & \AA\  & \AA\  & \AA\  & mag & mag & \AA\  & \AA\  & \AA\  &\AA\  & \AA\  & \AA\  \\
 \hline
-13.6& 0.847& 0.575& 6.912& 0.134& 0.219& 3.295& 2.453& 0.978& 1.331& 0.298&-0.252\\& 0.513& 0.198& 0.399& 0.006& 0.005& 0.197& 0.444& 0.254& 0.187& 0.164& 0.147\\
 -7.5& 4.809& 1.212& 4.718& 0.135& 0.234& 3.643& 2.108& 2.411& 1.470& 1.003& 0.384\\& 0.487& 0.188& 0.392& 0.006& 0.005& 0.188& 0.441& 0.239& 0.172& 0.159& 0.138\\
 -4.4& 5.452& 1.515& 5.284& 0.131& 0.251& 4.167& 2.717& 2.398& 1.209& 0.662& 0.629\\& 0.404& 0.155& 0.316& 0.005& 0.004& 0.153& 0.425& 0.199& 0.139& 0.139& 0.114\\
\end{tabular} 
\end{center}
\end{table*}

\clearpage

\section[]{Derived Aperture Data}
\label{app_apder}

Derived recession velocity, velocity dispersion, age, metallicity,
[Z/H], and $\alpha$-element enhancement, [E/Fe], and their $1\sigma$
errors for each aperture for each galaxy.

\begin{table*}
\begin{center}
\caption{Derived parameters and errors per aperture for NGC 3557.  The apertures from +8 -- +45$^{\prime\prime}$ are contaminated by another galaxy in the slit but the data are included here for completeness.}
\begin{tabular}{|rlclrl|}
\hline
Radius & RV & $\sigma$ & Log(Age) & $[$Z/H$]$ & $[$E/Fe$]$ \\
($^{\prime\prime}$) & (kms$^{-1}$) & (kms$^{-1}$) & Gyrs & dex & dex \\
 \hline
-45.04& 2929$\pm$  8& 36$\pm$23&0.550$\pm$0.091&-0.100$\pm$0.055&0.270$\pm$0.065\\
-36.13& 2924$\pm$  6& 63$\pm$18&1.175$\pm$0.028&-0.350$\pm$0.027&0.400$\pm$0.053\\
-30.88& 2924$\pm$  8& 73$\pm$15&1.175$\pm$0.028&-0.275$\pm$0.030&0.400$\pm$0.060\\
-26.86& 2921$\pm$  7& 89$\pm$18&1.175$\pm$0.032&-0.250$\pm$0.028&0.320$\pm$0.057\\
-23.65& 2925$\pm$  2& 99$\pm$17&1.175$\pm$0.007&-0.225$\pm$0.025&0.420$\pm$0.053\\
-21.09& 2918$\pm$  5&107$\pm$12&1.175$\pm$0.041&-0.200$\pm$0.033&0.360$\pm$0.060\\
\end{tabular} 
\end{center}
\end{table*}

\begin{table*}
\begin{center}
\caption{Derived parameters and errors per aperture for NGC 3640.}
\begin{tabular}{|rlclrl|}
\hline
Radius & RV & $\sigma$ & Log(Age) & $[$Z/H$]$ & $[$E/Fe$]$ \\
($^{\prime\prime}$) & (kms$^{-1}$) & (kms$^{-1}$) & Gyrs & dex & dex \\
 \hline
-44.9& 1199.$\pm$ 10.&139.$\pm$ 26.&0.850$\pm$0.080&-0.050$\pm$0.039&0.270$\pm$0.048\\
-31.5& 1207.$\pm$ 10.&131.$\pm$ 12.&0.950$\pm$0.068&-0.175$\pm$0.033&0.270$\pm$0.048\\
-25.5& 1212.$\pm$  4.&121.$\pm$ 21.&0.950$\pm$0.080&-0.125$\pm$0.038&0.240$\pm$0.046\\
-21.5& 1214.$\pm$  8.&125.$\pm$ 14.&0.950$\pm$0.070&-0.125$\pm$0.041&0.180$\pm$0.052\\
-18.7& 1216.$\pm$  6.&115.$\pm$ 15.&0.950$\pm$0.067&-0.150$\pm$0.045&0.270$\pm$0.058\\
-16.4& 1227.$\pm$  4.&119.$\pm$ 17.&0.875$\pm$0.080&-0.100$\pm$0.039&0.180$\pm$0.042\\
\end{tabular} 
\end{center}
\end{table*}

\begin{table*}
\begin{center}
\caption{Derived parameters and errors per aperture for NGC 5044.}
\begin{tabular}{|rlclrl|}
\hline
Radius & RV & $\sigma$ & Log(Age) & $[$Z/H$]$ & $[$E/Fe$]$ \\
($^{\prime\prime}$) & (kms$^{-1}$) & (kms$^{-1}$) & Gyrs & dex & dex \\
 \hline
-49.6& 2792.$\pm$ 14.&196.$\pm$ 21.&0.250$\pm$0.026& 0.800$\pm$0.095&0.150$\pm$0.037\\
-42.8& 2805.$\pm$ 12.&225.$\pm$ 17.&0.950$\pm$0.078& 0.175$\pm$0.052&0.400$\pm$0.050\\
-38.2& 2779.$\pm$ 14.&198.$\pm$ 16.&0.875$\pm$0.093& 0.100$\pm$0.071&0.500$\pm$0.055\\
-34.5& 2802.$\pm$ 11.&201.$\pm$ 12.&1.125$\pm$0.013&-0.075$\pm$0.042&0.360$\pm$0.056\\
-31.5& 2801.$\pm$ 12.&200.$\pm$ 12.&1.175$\pm$0.008& 0.075$\pm$0.030&0.480$\pm$0.052\\
-29.0& 2791.$\pm$  8.&204.$\pm$ 10.&1.075$\pm$0.042& 0.100$\pm$0.033&0.380$\pm$0.043\\
\end{tabular} 
\end{center}
\end{table*}

\begin{table*}
\begin{center}
\caption{Derived parameters and errors per aperture for A754\#1.}
\begin{tabular}{|rlclrl|}
\hline
Radius & RV & $\sigma$ & Log(Age) & $[$Z/H$]$ & $[$E/Fe$]$ \\
($^{\prime\prime}$) & (kms$^{-1}$) & (kms$^{-1}$) & Gyrs & dex & dex \\
 \hline
-20.8&16512.$\pm$ 46.&256.$\pm$100.&0.800$\pm$0.128&-0.275$\pm$0.087&0.360$\pm$0.088\\
-10.7&16514.$\pm$ 14.&143.$\pm$ 37.&1.075$\pm$0.023&-0.330$\pm$0.032&0.460$\pm$0.063\\
 -6.5&16497.$\pm$ 21.&202.$\pm$ 22.&1.175$\pm$0.015&-0.225$\pm$0.022&0.180$\pm$0.040\\
 -4.2&16468.$\pm$ 19.&274.$\pm$  5.&1.175$\pm$0.000& 0.000$\pm$0.023&0.180$\pm$0.038\\
 -2.7&16466.$\pm$ 11.&235.$\pm$  7.&1.175$\pm$0.000& 0.050$\pm$0.025&0.270$\pm$0.035\\
 -1.6&16461.$\pm$ 13.&271.$\pm$  5.&1.175$\pm$0.000& 0.125$\pm$0.025&0.240$\pm$0.033\\
 -0.6&16478.$\pm$ 11.&295.$\pm$  9.&1.175$\pm$0.000& 0.200$\pm$0.015&0.270$\pm$0.025\\
  0.4&16463.$\pm$ 12.&312.$\pm$ 12.&1.175$\pm$0.005& 0.250$\pm$0.015&0.270$\pm$0.025\\
  1.5&16442.$\pm$ 15.&277.$\pm$  6.&1.175$\pm$0.008& 0.175$\pm$0.022&0.210$\pm$0.031\\
  2.6&16470.$\pm$ 18.&254.$\pm$ 15.&1.125$\pm$0.012& 0.175$\pm$0.028&0.210$\pm$0.030\\
  4.0&16444.$\pm$ 22.&208.$\pm$ 16.&1.175$\pm$0.008&-0.075$\pm$0.029&0.270$\pm$0.036\\
  6.5&16453.$\pm$ 37.&254.$\pm$  9.&0.950$\pm$0.078& 0.100$\pm$0.057&0.240$\pm$0.055\\
 10.9&16450.$\pm$ 16.&249.$\pm$ 45.&0.925$\pm$0.087&-0.200$\pm$0.051&0.210$\pm$0.079\\
 20.9&16448.$\pm$ 33.&162.$\pm$ 30.&1.175$\pm$0.074&-0.500$\pm$0.072&0.240$\pm$0.076\\
\hline
\label{appendix_param_a754}
\end{tabular} 
\end{center}
\end{table*}

\begin{table*}
\begin{center}
\caption{Derived parameters and errors per aperture for A970\#1.}
\begin{tabular}{|rlclrl|}
\hline
Radius & RV & $\sigma$ & Log(Age) & $[$Z/H$]$ & $[$E/Fe$]$ \\
($^{\prime\prime}$) & (kms$^{-1}$) & (kms$^{-1}$) & Gyrs & dex & dex \\
 \hline
 -7.2&17595.$\pm$ 25.&244.$\pm$118.&1.175$\pm$0.097&-0.500$\pm$0.081&0.300$\pm$0.095\\
 -4.2&17578.$\pm$ 16.&188.$\pm$ 11.&1.150$\pm$0.051&-0.150$\pm$0.035&0.270$\pm$0.062\\
 -2.6&17536.$\pm$ 14.&208.$\pm$ 14.&1.125$\pm$0.027&-0.050$\pm$0.039&0.340$\pm$0.065\\
 -1.6&17558.$\pm$ 16.&229.$\pm$ 13.&1.175$\pm$0.004& 0.050$\pm$0.026&0.270$\pm$0.036\\
 -0.6&17522.$\pm$ 10.&246.$\pm$  7.&1.175$\pm$0.004& 0.175$\pm$0.023&0.300$\pm$0.034\\
  0.4&17517.$\pm$  9.&274.$\pm$ 14.&1.175$\pm$0.000& 0.250$\pm$0.017&0.270$\pm$0.030\\
  1.5&17520.$\pm$  4.&256.$\pm$ 14.&1.125$\pm$0.009& 0.125$\pm$0.022&0.240$\pm$0.033\\
  2.5&17557.$\pm$ 12.&255.$\pm$ 18.&1.125$\pm$0.007& 0.000$\pm$0.026&0.360$\pm$0.051\\
  4.0&17528.$\pm$ 18.&183.$\pm$ 30.&1.025$\pm$0.034&-0.050$\pm$0.032&0.240$\pm$0.055\\
  7.1&17573.$\pm$ 36.&256.$\pm$ 27.&1.175$\pm$0.031&-0.300$\pm$0.032&0.300$\pm$0.071\\
\hline
\label{appendix_param_a970}
\end{tabular} 
\end{center}
\end{table*}

\begin{table*}
\begin{center}
\caption{Derived parameters and errors per aperture for A978\#1.}
\begin{tabular}{|rlclrl|}
\hline
Radius & RV & $\sigma$ & Log(Age) & $[$Z/H$]$ & $[$E/Fe$]$ \\
($^{\prime\prime}$) & (kms$^{-1}$) & (kms$^{-1}$) & Gyrs & dex & dex \\
 \hline
-13.6&17736.$\pm$ 29.&233.$\pm$ 55.&1.175$\pm$0.063&-0.275$\pm$0.038&0.240$\pm$0.068\\
 -7.5&17714.$\pm$ 11.&242.$\pm$ 33.&1.175$\pm$0.011&-0.175$\pm$0.020&0.150$\pm$0.044\\
 -4.4&17729.$\pm$ 13.&183.$\pm$  9.&1.175$\pm$0.000&-0.100$\pm$0.017&0.210$\pm$0.037\\
 -2.6&17724.$\pm$ 14.&269.$\pm$ 20.&1.025$\pm$0.031& 0.075$\pm$0.029&0.150$\pm$0.031\\
 -1.6&17688.$\pm$ 13.&285.$\pm$ 15.&1.100$\pm$0.008& 0.000$\pm$0.024&0.030$\pm$0.042\\
 -0.6&17669.$\pm$ 14.&308.$\pm$ 15.&1.025$\pm$0.025& 0.125$\pm$0.032&-.060$\pm$0.034\\
  0.4&17618.$\pm$ 10.&301.$\pm$ 12.&1.000$\pm$0.031& 0.175$\pm$0.037&0.000$\pm$0.043\\
  1.5&17632.$\pm$ 14.&247.$\pm$ 12.&1.175$\pm$0.000& 0.000$\pm$0.014&0.210$\pm$0.031\\
  2.5&17601.$\pm$ 10.&237.$\pm$ 21.&1.175$\pm$0.007&-0.025$\pm$0.027&0.120$\pm$0.037\\
  4.2&17607.$\pm$ 17.&253.$\pm$ 22.&1.175$\pm$0.004&-0.100$\pm$0.021&0.210$\pm$0.037\\
  7.4&17581.$\pm$ 19.&248.$\pm$ 32.&1.175$\pm$0.007&-0.200$\pm$0.031&0.060$\pm$0.064\\
 13.4&17697.$\pm$ 42.&264.$\pm$ 23.&1.175$\pm$0.020&-0.075$\pm$0.026&0.400$\pm$0.049\\
\hline
\label{appendix_param_a978}
\end{tabular} 
\end{center}
\end{table*}



\bsp

\label{lastpage}

\end{document}